\pdfoutput=1 
\documentclass[aps,prd,amsmath,floats,floatfix,onecolumn,superscriptaddress,nofootinbib,showpacs]{revtex4-2}

\usepackage[T1]{fontenc}
\usepackage[utf8]{inputenc}
\usepackage{lmodern}
\usepackage{lipsum}
\usepackage{phoenician}
\usepackage[english]{babel}

\usepackage[sort&compress]{natbib}
\usepackage{caption}

\usepackage[dvipsnames, usenames]{xcolor}
\usepackage{colortbl}

\usepackage[dvipsnames, usenames]{xcolor}
\usepackage{colortbl}
\definecolor{linkcolor}{rgb}{0.0,0.3,0.5}
\usepackage[hypertexnames=false, unicode, colorlinks=true, linkcolor=linkcolor,
citecolor=linkcolor, filecolor=linkcolor,urlcolor=linkcolor,
pdfusetitle]{hyperref}

\usepackage[pdftex]{graphicx}
\usepackage{xspace}
\usepackage{appendix}
\usepackage{bm}

\usepackage{physics}
\usepackage{ragged2e}

\usepackage{microtype}

\usepackage[english]{babel}
\usepackage{blindtext}

\DeclareMathAlphabet{\mathpzc}{OT1}{pzc}{m}{it}

\newcommand*{\fnl} {f_{\rm NL}}

\usepackage{comment}
\usepackage[thinc]{esdiff}
\newcommand{\dif}{\mathrm{d}}

\definecolor{darkred}{RGB}{175,0,0}
\definecolor{darkblue}{RGB}{14,0,185}
\definecolor{salmon}{RGB}{255,160,105}

\begin{document}
\title{Full-sky spherical Fourier-Bessel power spectrum in general relativity}

\newcommand\federicohome{
\affiliation{Dipartimento di Fisica Galileo Galilei, Universit\` a di Padova, I-35131 Padova, Italy}
\affiliation{INFN Sezione di Padova, I-35131 Padova, Italy}
}

\newcommand\danielehome{
\affiliation{Dipartimento di Fisica Galileo Galilei, Universit\` a di Padova, I-35131 Padova, Italy}
\affiliation{INFN Sezione di Padova, I-35131 Padova, Italy}
\affiliation{INAF-Osservatorio Astronomico di Padova, Italy}
}

\newcommand\alvisehome{
\affiliation{Dipartimento di Fisica Galileo Galilei, Universit\` a di Padova, I-35131 Padova, Italy}
\affiliation{INFN Sezione di Padova, I-35131 Padova, Italy}
\affiliation{INAF-Osservatorio Astronomico di Padova, Italy}
}

\author{Federico Semenzato}
\email{federico.semenzato.1@phd.unipd.it}
\federicohome

\author{Daniele Bertacca}%
\danielehome

\author{Alvise Raccanelli}%
\alvisehome

\date{\today}

\begin{abstract}
We present a formalism for analyzing galaxy clustering on the lightcone with the two-point correlation in the spherical Fourier-Bessel (SFB) formalism, which is a natural choice to account for all wide-angle and relativistic (general relativity, GR) effects.
We extend previous studies by including all projection and GR effects, developing an efficient numerical implementation that avoids the use of the Limber approximation, includes multibins correlations and a full nondiagonal covariance.

Using this formalism, we investigate the impact
of neglecting GR corrections, and in particular how much this could bias measurements of the non-Gaussianity parameter $\fnl$.
Our results show that not including relativistic projection terms can systematically and non-negligibly bias estimates of $\fnl$. The exact results depend on survey specifications and galaxy population properties, but we stress that a bias will generally be present.

Finally, we develop a novel prescription for cross-bin correlations that allow to search for a clean signal of relativistic corrections, and show that this requires the use of the 3D full-sky formalism.
\end{abstract}

\maketitle

\section{Introduction}
\label{sec:intro}
Large-scale galaxy surveys provide maps of three-dimensional positions of millions of galaxies, with the next generation of surveys further increasing the scale and precision of such maps, effectively resembling a sort of cosmological cartography.
These datasets will allow to advance our understanding of the Universe, putting tighter constraints on cosmological models. The well-established Lambda Cold Dark Matter ($\Lambda$CDM) model still lacks solid theoretical foundations, leaving us to hunt for the mechanism of the current cosmic acceleration~\cite{Riess_1998, Perlmutter_1999, Guz+08}, the nature of dark matter and dark energy \cite{Hinshaw_2013,planck,Bird_2016} and testing the validity of general relativity (GR)~\cite{Barausse_2005,Wang_2008,DVALI2000208,Ishak_2018}. The statistical analysis of galaxy clustering represents a fruitful research field~\cite{Tegmark_1997, Verde_1998, Percival_2004, Weinberg_2004,Verde_2010, Huterer_2015,Mueller_2018, desicollaboration2024desi}, and as the amount of available data increases and larger volumes are probed, it is crucial to properly model the observable quantities to construct robust predictions and to maximize the information extracted by these new datasets~\cite{Bonvin11,Bertacca_2ndorder,Raccanelli_2016,beware1}.

Since past surveys covered narrow regions of the sky, the so-called plane parallel approximation has been implemented~\cite{kaiser}: in this, the curvature of the observed sky is neglected and sources are placed at the same distance from the observer, and it is designed for linear scales.
This approach is valid for past smaller surveys. However, this is no longer an accurate approximation for future surveys (such as e.g.,~SPHEREx~\cite{spherex_cosmo}, Euclid~\cite{euclid_report,euclidcollaboration2024euclid}, ATLAS Probe~\cite{wang_atlas_2}, DESI~\cite{desicollaboration2016desi,desicollaboration2024desi}, Nancy Grace Roman Space Telescope~\cite{spergel2015widefield}, Megamapper~\cite{schlegel2022megamapper}), as they will span large volumes and high fractions of the sky~\cite{Szalay_1998,Bharadwaj_1999,Matsubara_2000,Bertacca_2012,YD13,Raccanelli_2015,fftlog,theotherFS}. Moreover, observations actually probe our past light cone; thus, the light path between the source and the observer is affected by additional projection and distortion effects in the measured quantities (refer to Fig. 1 from~\cite{Raccanelli_2016} for a visual summary) that must be properly taken into account in the analysis~\cite{ Matsubara_1997, Yoo_2009, Bonvin11, Challinor_2011, Jeong_2012, bertacca_beyond_2012, Dio_2013, Raccanelli_2016_2, Alonso_2015, Bonvin_2016, Dio_2016, Gaztanaga_2017, Bertacca_2020}.

Maintaining systematic biases below statistical errors  and accurately assessing true uncertainties are crucial tasks for precision cosmology analyses. The huge complexity and computational cost required to precisely map observable quantities to parametric models can be addressed by introducing careful approximations without sacrificing accuracy significantly. Still, inaccurate approximations in the analysis impact inferred model parameters by shifting the posterior distribution peak and distorting its shape, leading to errors in parameter values and their error bars~\cite{raccanelli_neutrinos,beware2}.

In the context of galaxy clustering, various summary statistics have been developed to extract cosmological information from galaxy surveys. The two-point correlation function power spectrum is the most common statistics~\cite{bible}. However, the standard Fourier power spectrum is not suited to analyze large angular correlations, where the assumption of a single line of sight must be dropped. This ``particles in a box'' power spectrum is not a proper observable if one wants to account for the presence of an observer and for the degeneracy of the radial coordinate with time. Moreover, in standard $C_\ell$ analyses, wide-angle correlations are naturally embedded while radial correlations are either neglected within a redshift bin, or require a large number of thin bins, making the full analysis very computationally expensive. This issue must also be addressed when dealing with relativistic effects, which are expected to be detectable in future surveys.

One approach that accommodates both radial and angular scales of forthcoming surveys is the spherical Fourier-Bessel (SFB) transform, firstly developed in~\cite{fish95,Heavens_1995}. This method provides an orthogonal basis for all-sky analyses operating in spherical coordinates, which naturally match angular/radial separations~\cite{rassat,pullen,sfb_analysis,Wang_2020}. SFB inherently considers the line of sight for each galaxy, enabling straightforward modeling of redshift evolution, galaxy bias, and growth factors. In~\cite{Percival_2004_sfb}, the SFB decomposition has been performed on the Two-degree-Field Galaxy Redshift Survey
(2dFGRS) galaxy overdensity field to extract constraints on cosmological parameters. In~\cite{YD13}, the authors developed the formalism for the SFB power spectrum including first order relativistic corrections, and in~\cite{wen2024exact}, the authors model wide-angle and Doppler effects in the SFB basis. 
A refined estimator for the spherical power spectrum has been developed and validated in~\cite{superfab,khek2022fast,doresfbvalidation}, including redshift-space distortions, primordial non-Gaussianity (PNG) and characteristic survey effects. 
A fully relativistic expansion up to second order is provided in~\cite{Bertacca_2018}, while a first numerical implementation of the (nonrelativistic) SFB bispectrum has been presented in~\cite{dorebispectrum}.

In this work, we present a formalism for the full-sky analysis of the SFB power spectrum in GR. We extend previous studies~\cite{Lanusse_2015,pullen} by including all projection effects, avoiding the use of the Limber approximation, and including multiwindows and off-diagonal terms of the covariance. We use our code to forecast (for a SPHEREx-like survey) detectability of relativistic effects.
Crucially, we show that using approximations such as flat sky and neglecting relativistic terms would bring 
systematic shifts in cosmological parameters and Primordial Non-Gaussianity PNG constraints. Furthermore, we provide a first analysis of cross-bin correlations in the SFB base to isolate the contributions of integrated relativistic effects.

The paper is organized as follows. In Sec.~\ref{sec:greff}, we discuss relativistic corrections to the observed galaxy number count and introduce our main observable. In Sec.~\ref{sec:sfbintro}, we briefly review  the SFB formalism, its application to galaxy clustering and the forecasting tools used in the analysys. In Sec.~\ref{sec:results} we present our forecasts for a SPHEREx-like survey and address the impact of GR effects on cosmological parameters.  In Sec.~\ref{sec:MW} we develop a multi-window approach to perform cross-bin correlations in SFB. Finally, in Sec.~\ref{sec:conclusions}, we draw our conclusions.

\section{Galaxy Clustering in General Relativity}
\label{sec:greff}
We can only observe galaxies in our past light cone. The presence of inhomogeneities in real space deflects the null geodesics of the photons emitted from a galaxy and propagating to the observer. This implies that observable quantities (i.e.,~angular position in the sky, redshift and photon flux at any given waveband) differ in the case of an unperturbed or perturbed Universe, and that 3D maps of galaxy distributions are distorted. These distortions are GR effects that arise on cosmological scales, both local and nonlocal (i.e.,~integrated along the line of sight), and they are commonly referred to as GR projection effects~\cite{Challinor_2011,Bonvin11,Jeong_2012,yoo_releff,raccanelli2016doppler, Bertacca_2ndorder}.

It is crucial to model the observed galaxy overdensity $\delta_{\mathrm{obs}}$, which is gauge invariant and unique. An overdensity $\delta_g$ defined in different gauges matches the observed one on sub-Hubble scales, but differences arise on large scales. Upcoming surveys will cover extremely vast regions of the sky, close to the size of the horizon, where GR contributions might not be disregarded. 

Large survey volumes allow for high-precision measurements of large-scale structure clustering. High accuracy is required when analyzing such high-precision data: refined theoretical modeling is of fundamental importance to properly interpret the measured clustering signal, and the impact of GR effects on statistics of galaxy distributions must be properly quantified and taken into account.

Moreover, the spatially averaged number density of galaxies, $\bar{n}_g(z)$, evolves in time since new galaxies form and hence the true number density does not scale like $a^{-3}$ (where $a=1/(1+z)$ is the cosmological
scale factor). Its observed value depends on the specifics of the survey and the underlying physics. We label the galaxy number density as $n_g(\mathbf{x})$ and define it as
\begin{equation}\label{numdens}
    n_g(\mathbf{x})= \bar{n}_g(z)\qty(1+\delta_g(z, \hat{\mathbf{x}}))\,,
\end{equation}
where $\delta_g(z, \hat{\mathbf{x}})$ is the galaxy overdensity field and the position vector $\mathbf{x}$ depends on the radial position $r(z)=\abs{\mathbf{x}}$ and angular position $\hat{\mathbf{x}}$ in the observed frame.

To account for survey geometry specifications, one can introduce observational selection and window functions, $\phi(\mathbf{x})$ and $W(\mathbf{x})$ respectively. It is very important to properly model them to perform the projection on the SFB basis. A window function can be implemented to exclude or isolate specific regions, so that 
\begin{equation}
    V_\text{survey}=\int_0^\infty W(\mathbf{x}) \,d^3\mathbf{x}\,.
\end{equation}
Window functions act on the entire density field, i.e.,~$n_g(\mathbf{x}) \xrightarrow[]{} n_g^W(\mathbf{x}) \equiv W(\mathbf{x})n_g(\mathbf{x})$, and can be related to the observed redshift distribution \(\mathcal{N}_z(z)\) and in turn to the survey volume $V_{\rm survey}$ as
\begin{equation}
\int d r\, r^2 W(r)=\int d z \frac{c}{H(z)} r^2 W(r)=\int d z \mathcal{N}_z(z)=\frac{V_\mathrm{survey}}{4 \pi }\,,
\end{equation}
where the observed galaxy position is decomposed in  $\mathbf{x}\equiv(r(z),\hat{\mathbf{x}})$ and normalized to \(\int dz\, \mathcal{N}_z(z)=1\). 
The complete treatment for an angular masked survey in SFB and its statistical implications are provided in~\cite{superfab}.

On the other hand, a selection function of the galaxy survey $\phi(z)$ acts on \( \bar{n}_g(z)\) as
\begin{equation}\label{nbarsel}
    \bar{n}_g(z)\equiv\Tilde{n}_g \phi(z)\equiv\frac{N_g^\text{obs}}{V_\text{survey}}\phi(z)\,,
\end{equation}
where $\Tilde{n}_g$ is the mean number density of galaxies in the survey and $N_g^\text{obs}$ is the total number of observed galaxies.

The galaxy number density field of Eq.~\eqref{numdens} then reads
\begin{equation}\label{ngx}
    n_g(\mathbf{x})=\tilde{n}_g \phi(z)\qty[1+\delta_g(z, \hat{\mathbf{x}})]\, .
\end{equation}

In this work, the power spectrum of the comoving curvature perturbations $\zeta(\mathbf{k})$  is written in the standard adimensional normalization as 
\begin{equation}
    \Delta_\zeta(k)=k^3 \frac{P_\zeta(k)}{2\pi^2}\,,
\end{equation}
where the power spectrum is almost scale invariant at initial epoch and the transfer function is angle independent.
One can then write the observed overdensity as
\begin{equation}\label{transfer_deltag}
    \delta_g(z, \hat{\mathbf{x}})=\mathcal{T}_g(k,z)\zeta(\mathbf{k})+\epsilon(\mathbf{k},z)\,,
\end{equation}
and $\epsilon(\mathbf{k},z)$ is a residual Poisson noise (arising from the discreteness of the galaxy distribution~\cite{Feldman_1994}), which is not correlated with $\delta_g(z, \hat{\mathbf{x}})$. The noise power spectrum reads 
\begin{equation}
    \langle\epsilon(\mathbf{k},z)\epsilon(\mathbf{k}',z)\rangle=\frac{1}{\tilde{n}_g \phi(z)}(2\pi)^3\delta^D(\mathbf{k}-\mathbf{k}')\,.
\end{equation}

\subsection{General relativistic effects}
\label{sec:greffects}

The observed galaxy number density is affected by the propagation of photons in an inhomogeneous universe. As introduced in the previous section, the observed redshift and angular position of a galaxy (i.e.,~the photons direction of propagation) are distorted with respect to the physical ones. The redshift measurement is affected by the peculiar velocity of the galaxy~\cite{kaiser}. Gravitational potential and lensing also introduce redshift and observed direction distortions, and the observed volume differs from the physical one. In addition, the perturbations of the metric and the matter field are gauge dependent, and the observed galaxy number count is a gauge-invariant quantity. All of these effects are encoded at first order in the observed galaxy number count $n_g^{\mathrm{obs}}(z,\mathbf{\hat{x}})$. In this section, we briefly review the main relativistic effects on the observed galaxy number count (see~\cite{Challinor_2011,Bertacca_2ndorder,Bertacca_2018, yoo_releff,YD13}). 

The observed number density of galaxies in the observed volume element $\mathrm{d}V_{\mathrm{obs}}$  is computed at redshift $z$ and angular position $\mathbf{\hat{x}}$ by counting the number of galaxies $\mathrm{d}N_g(z,\mathbf{\hat{x}}) $ within the observed volume, i.e.,
\begin{equation}
    n_g^{\mathrm{obs }}(z,\mathbf{\hat{x}})=\diff{N_g(z,\mathbf{\hat{x}})}{V_{\mathrm{obs}}}\,.
\end{equation}
Here, the observed volume is defined at the observed redshift $z$ and galaxy position $\{\theta,\phi\}$:
\begin{equation}
\mathrm{d} V_{\mathrm{obs}}(z)=\frac{r^2(z)}{H(z)} \sin \theta \, \mathrm{d} z \mathrm{d} \theta \mathrm{d} \phi\,.
\end{equation}
The galaxy number overdensity $\delta_g(z, \mathbf{\hat{x}})$ is then defined as
\begin{equation}
n_{g}^{\mathrm{obs}}(z, \mathbf{\hat{x}}) \equiv \bar{n}_{g}(z)\qty(1+\delta_g(z, \mathbf{\hat{x}})) \,,
\end{equation}
where $\bar{n}_{g}(z) \equiv\langle n_{g}^{\rm obs }(z)\rangle$ is the observed galaxy number density averaged over the solid angle at fixed redshift and $\langle\delta_g\rangle=0$. Note that this holds for full-sky surveys and should be properly adjusted for a sky-fraction coverage $f_{\rm sky} <1$~\cite{Challinor_2011}, which implies additional correlations between $\ell$ modes~\cite{superfab}. 
The total observed overdensity $\delta_g^{\mathrm{obs}}(z,\mathbf{\hat{x}})$ is expressed as a combination of terms that encode both the local density field  (i.e., effects sourced at the galaxy location)  and contributions integrated along the line of sight along the photon propagation path. The observed overdensity then reads~\cite{surveyspecs}
\begin{equation}\label{observed_deltag}
    \delta_{\mathrm{obs}}(\mathbf{\hat{x}}, z)=\delta_g^{\mathrm{int}}(\mathbf{\hat{x}}, z)+\delta_{\mathrm{rsd}}(\mathbf{\hat{x}}, z)+\delta_{\mathrm{v}}(\mathbf{\hat{x}}, z)+\delta_\kappa(\mathbf{\hat{x}}, z)+\delta_{\mathrm{pot}}(\mathbf{\hat{x}}, z)
    \end{equation}
where $\delta_g^{\mathrm{int}}$ is the intrinsic galaxy overdensity in the comoving gauge, $\delta_{\mathrm{rsd}}$ encode peculiar velocity contributions, i.e. redshift-space distortions in the Kaiser approximation, $\delta_{\mathrm{v}}$ accounts for Doppler effects, $\delta_\kappa$ is the lensing convergence and $\delta_{\mathrm{pot}}$ is written as a function of Bardeen potentials and their temporal evolution. In Appendix~\ref{app:GR_def}, we provide the definitions of such terms.
Lensing convergence $\kappa$ results in changes to the apparent angular position, while the gravitational potential term affect the observed radial position. A Doppler lensing term is introduced to encode apparent changes in the angular size and magnitude of a galaxy caused by the peculiar motion of the source with respect to the observer. This effect arises from the aberration effect in special relativity~\cite{Bacon_2014}.
A time delay and integrated Sachs-Wolf terms arise from the gravitational potential contributions, which directly involve Bardeen potentials~\cite{surveyspecs}.

\section{Spherical Fourier-Bessel Power Spectrum}
\label{sec:sfbintro}
In this section, we briefly introduce and review the spherical Fourier decomposition~\cite{fish95,Heavens_1995}.  

The complete radial and angular basis in a spherical Fourier-Bessel space $\ket{k \ell m}$ is defined as~\cite{Bertacca_2018,superfab}
\begin{equation}
\label{eq:klm_def}
\braket{\mathbf{x}}{k \ell m}=\sqrt{\frac{2}{\pi}}k  j_{\ell}(k r) Y_{\ell m}(\hat{\mathbf{x}})\,,
\end{equation}
where $\mathbf{x}=r\hat{\mathbf{x}}$ is the position vector and $\mathbf{k}=k\hat{\mathbf{k}}$ encodes the Fourier mode, with $r=\abs{\mathbf{x}}$, $k=\abs{\mathbf{k}}$ and $\hat{\mathbf{x}}$, $\hat{\mathbf{k}}$ unit directional vectors. $Y_{\ell m}(\hat{\mathbf{x}})$ is a spherical harmonic and $j_{\ell}(k r)$ a spherical Bessel function.

With this prescription, any scalar field $\varphi=\ket{\varphi}$ can be projected to the $\ket{k\ell m}$ basis through
\begin{equation}
    \braket{k \ell m}{\varphi}=\varphi_{\ell m}(k)\,.
\end{equation}
The coefficients are fixed to match the normalization employed in~\cite{YD13}, but the derivation can be performed in full generality as in~\cite{Bertacca_2018}.

For a generic scalar field $\varphi$, the spherical power spectrum $\mathcal{S}_\ell(k,k')$ is defined as the ensemble average of the product of the spherical Fourier modes of the field $\varphi_{\ell m}(k)$ and its complex conjugate $\varphi_{\ell' m'}^*(k')$. That is~\cite{rassat,YD13},
\begin{equation}
\langle\varphi_{\ell m}(k) \varphi_{\ell' m'}^*(k')\rangle \equiv \delta_{\ell \ell'} \delta_{m m'} \mathcal{S}_\ell(k, k')\,,
\end{equation}
where $\delta_{\ell \ell'}$ and $\delta_{m m'}$ are Kronecker delta functions. Different angular modes are uncorrelated if the scalar field is defined at every angular direction (i.e.,~no angular mask is applied). The standard power spectrum  $P(\mathbf{k}, \mathbf{k}')$ is related to the Fourier modes of the field and is defined as the ensemble average
\begin{equation}\label{reducestopk}
\langle\varphi(\mathbf{k}) \varphi^*(\mathbf{k}')\rangle \equiv P(\mathbf{k}, \mathbf{k}')=(2 \pi)^3 \delta^D(\mathbf{k}-\mathbf{k}') P(k)\,,
\end{equation}
where the Dirac delta function $\delta^D(\mathbf{k}-\mathbf{k}')$ enforces statistical isotropy. Hence, for a rotationally and translationally invariant power spectrum, the spherical one reduces to (see e.g.,~\cite{raccanelli2023power,Gao_2023})
\begin{equation}\label{Sl_to_Pk}
     \mathcal{S}_\ell(k, k')=\delta^D(k-k')\mathcal{S}_\ell(k)=\delta^D(k-k')P(k)\,,
\end{equation}
matching the spherical prescription to the standard Fourier one. 

Applying this  formalism to the overdensity field, one can then retrieve its spherical power spectrum. In the following discussion, we will consider a full-sky survey and embed its radial features inside the selection function $\phi(r)$, while the window function $W(r)$ is set as constant. In Sec.~\ref{sec:MW} we extend the treatment to correlations between different radial bins, which will be included through different $W_1$, $W_2$. In Appendix~\ref{app:coeff}, we provide more details about the SFB basis and sketch the derivation of the two-point correlation function $\xi_g\qty(\mathbf{x}_1-\mathbf{x}_2)$ and the spherical power spectrum $\mathcal{S}_\ell(k,k')$ for the observed galaxy overdensity field in redshift space~\cite{Feldman_1994,YD13}.

We write the spherical Fourier power spectrum as the sum of two contributions, a signal one $\bar{\mathcal{S}}_\ell\qty(k, k')$ and a \textit{noise} one $\mathcal{N}_\ell\qty(k, k')$ i.e.,
\begin{equation}\label{sig+noise}
\mathcal{S}_\ell\qty(k, k') \equiv \bar{\mathcal{S}}_\ell\qty(k, k')+\mathcal{N}_\ell\qty(k, k')\,.
\end{equation}
The signal component i.e., the $\xi_g$-dependent term in Eq.~\eqref{sl_2point} reads
\begin{equation}\label{signal_def}
\mathcal{\bar{S}}_\ell(k, k')=4 \pi \int d \ln \tilde{k} \, \Delta_{\zeta}(\tilde{k}) \,\mathcal{M}_\ell(\tilde{k}, k) \,\mathcal{M}_\ell(\tilde{k}, k')\,,
\end{equation}
where expression is made more compact by introducing the spherical multipole function $\mathcal{M}_\ell(\tilde{k}, k)$, that is
\begin{equation}\label{sphmult}
\mathcal{M}_\ell(\tilde{k}, k) \equiv k \sqrt{\frac{2}{\pi}} \int_0^{\infty} d r\, r^2\, W(r)\,\phi(r) \,j_\ell(\tilde{k} r) \,j_\ell(k r) \,\mathcal{T}_g(\tilde{k}, r)\,.
\end{equation}
A transfer function $\mathcal{T}_g$ connects the galaxy power spectrum to the standard power spectrum of primordial curvature perturbation $\zeta$.

On the other hand, the shot-noise term which reads (refer to Appendix~\ref{app:coeff} for further details)
\begin{equation}\label{noise_def}
\mathcal{N}_\ell\qty(k, k') \equiv \frac{2 k k'}{\pi \tilde{n}_g} \int_0^{\infty} dr\, r^2\, W^2(r)\,\phi(r) \,j_\ell(k r) \,j_\ell\qty(k' r)\,.
\end{equation} 

All terms $\delta_i$ in Eq.~\eqref{observed_deltag} can be related back to the matter density component $\delta_m$ and its transfer function $\mathcal{T}_m(k,r)$~\cite{Eisenstein_1998} (normalized at $z=0$) through specific weight functions $\Xi_i(\mathbf{x},\mathbf{k})$, so that~\cite{YD13}
\begin{equation}\label{GR_eff}
\delta_i^{\mathrm{obs}}(\mathbf{x})=\int_0^r d \tilde{r} \int \frac{d^3 \mathbf{k}}{(2 \pi)^3} \Xi_i(\mathbf{x},\tilde{\mathbf{x}}, \mathbf{k}) \mathcal{T}_m(k, \tilde{r}) \zeta(\mathbf{k}) e^{i \mathbf{k} \cdot \tilde{\mathbf{x}}} + \epsilon(\mathbf{x})\,,
\end{equation}
where $\tilde{\mathbf{x}}=(\tilde{r},\hat{\mathbf{x}})$ encodes the line-of-sight position in spherical coordinates and $\epsilon(\mathbf{x})$ is the residual shot noise contribution. 
The expansion can now be performed in this framework, producing an updated structure for the spherical multipole $\mathcal{M}_\ell^i(\tilde{k},\tilde{k})$
\begin{equation}
\label{sph_mult_win}
\mathcal{M}_\ell^i(\tilde{k}, k) \equiv k \sqrt{\frac{2}{\pi}} \int_0^{\infty} d r \, r^2\, W(r)\,\phi(r) \, j_{\ell}(k r)\,
\int_0^{r} d\tilde{r} \,\Xi^i_\ell(r,\tilde{r},\tilde{k})\,\mathcal{T}_m(\tilde{k}, \tilde{r})\,j_\ell(\tilde{k} \tilde{r})\,,
\end{equation}
where $\Xi^i_\ell(r,\tilde{r},\tilde{k})$ is the expansion of the weight function. The multipoles can then be properly summed into a single $\mathcal{M}_\ell(\tilde{k},k)=\sum_i q_i\mathcal{M}^i_\ell(\tilde{k},k)$, where $q_i$ are the distance-independent numerical coefficients associated to each effect. The full expressions for the weight functions are reported in~\cite{YD13,Bertacca_2018}. As previously stated in Eq.~\eqref{Sl_to_Pk}, for single contributions with isotropic power spectrum, their spherical power spectrum will almost match the three-dimensional one $\mathcal{S}^i_\ell(k)\equiv\mathcal{S}^i_\ell(k,k)\simeq P(k)$, with deviations explained in the following paragraphs. 
Appendix~\ref{app:numint} includes some insights on the numerical implementation of the spherical Bessel nested integrals. 

\begin{figure}
    \centering
    \includegraphics[width=\textwidth]{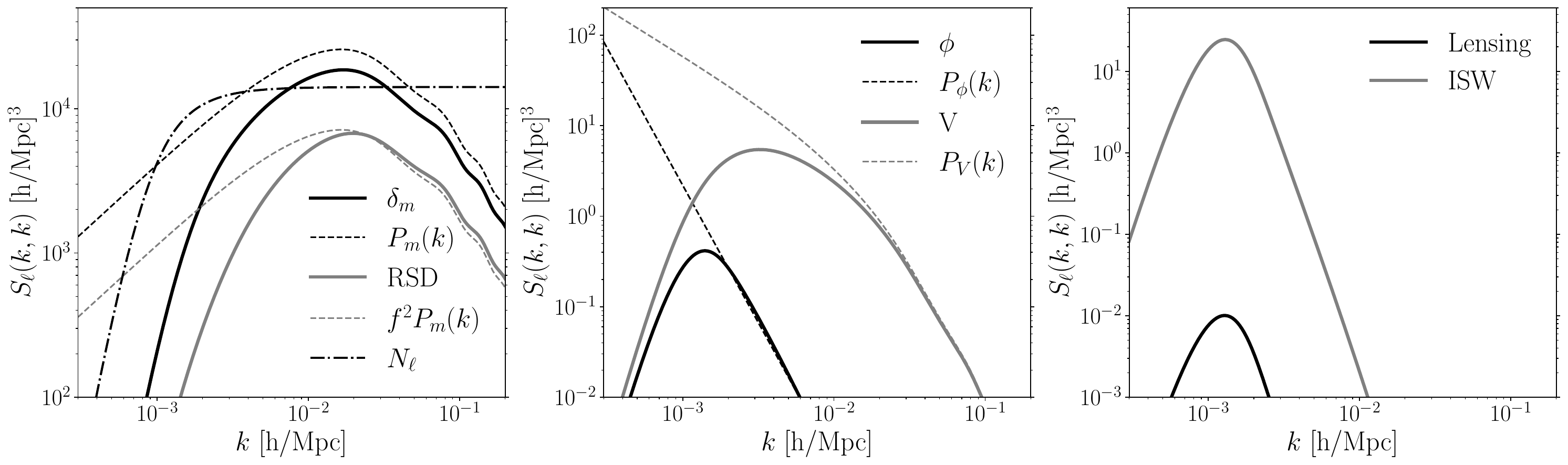}
    \caption{\justifying Spherical power spectra for a reference Gaussian selection function case assuming  a constant $\tilde{n}_g=10^{-4}(h^{-1} \mathrm{Mpc})^{-3}$. Different GR contributions are plotted against their corresponding three-dimensional quantity for comparison. The shot noise is shown in black in the left panel and it can be seen how it saturates to a constant value at small scales as expected.}
    \label{fig:sfbintro}
\end{figure}

As a first benchmark case, the spherical power spectrum $\mathcal{S}_\ell(k,k')$ is computed for a survey modeled with a Gaussian selection function
\begin{equation}
\label{eq:gaussprova}
    \phi_{G}(r)=\exp \qty[-\qty(\frac{r}{r_0})^2]\,,
\end{equation}
where $z_0(r_0) = 1$. The left panel of Fig.~\ref{fig:sfbintro} shows the diagonal spherical power spectra $\mathcal{S}_\ell(k,k)$ of the matter density $\delta_m$, the redshift-space distortion (RSD) term and the shot noise  $\mathcal{N}_\ell(k,k)$ at $\ell=2$. In the Figure, the spherical power spectrum and shot noise are normalized as in~\cite{YD13} by computing their small-scale limit for the specific choice template selection function. In the following sections, the power spectrum is computed in full generality for all the selection and window functions employed.

For comparison, the matter density power spectrum $P_m(k)$ is also plotted.  Note that since $P_m(k)$ is isotropic, the matter density contribution  $\mathcal{S}^{\delta_m}_\ell(k,k)$ is independent of angular multipole $\ell$ and at small scales $P_m(k)\sim \mathcal{S}^\delta_\ell(k,k)$ is almost retrieved, although the power is damped for the $\delta_m$ component. The mismatch  is due to the redshift evolution of the integrated quantities in Eq.~\eqref{sphmult}, since the averaging is done over a range of redshifts. 

On the other hand, the RSD component has a slight increase in amplitude with respect to its three-dimensional counterpart $P_\mathrm{RSD}(k,\mu_k=1)$ (where $\mu_k=\hat{\mathbf{x}}\cdot\hat{\mathbf{k}}$ and $\mathbf{x}$ is the single line of sight direction). This is due to the integration of the weight function defined as (cf. Eq.~\eqref{GR_eff})
\begin{equation}
    \Xi_{\ell}^{\mathrm{RSD}}(r,\tilde{r},\tilde{k})=-\delta^D(r-\tilde{r})\qty[f(r)\frac{\partial^2}{(\tilde{k}\partial r)^2}]\,.
\end{equation}
Here, the logarithmic growth rate of structure $f(z)=d \ln{D(z)}/d\ln{a}$ (with $a$ the cosmological scale factor $a=1/(1+z)$) is higher for higher $z$, thus inducing an increase with respect to the single-redshift three-dimensional counterpart. 

The survey selection function approaches zero for radial distances $r\gtrsim r_0$, making it that the spherical power spectrum $\mathcal{S}_\ell(k,k)$ is highly suppressed. The actual largest scale a survey can probe is discussed is Sec.~\ref{subsec:modsel},
and Appendix~\ref{app:kdisc} discusses the discretization of the spherical power spectrum and the choice of the radial binning for a finite survey.

The central plot of Fig.~\ref{fig:sfbintro} shows the contributions of the line-of-sight velocity $V$ and the gravitational potential $\phi$. For comparison, their three-dimensional power spectra are also plotted.
Although $P_\phi(k)$ is independent of multipole $\ell$, $\mathcal{S}_\ell^\phi(k,k)$ drops in amplitude as $\ell$ increases at fixed $k$, due to the finite-survey suppression at $k\leq k_c(\equiv \ell/r_0)$ moving at smaller scales (higher $k$) for higher $\ell$s.

The line-of-sight velocity spherical power spectrum $\mathcal{S}_\ell^V(k)$ is also $\ell$ dependent and decreasing for higher $\ell$. The velocity and potential contributions, respectively, scale as $(k/\mathcal{H})$ and $(k/\mathcal{H})^2$ relative to the $\delta_m$ contribution~\cite{Bertacca_2ndorder,YD13}. This implies that they will not be relevant in small galaxy surveys, being comparable to the matter density term only at horizon scale $k\simeq \mathcal{H}$. On the other hand, they become relevant for wide-field surveys and they are included in this analysis to provide the full picture of all first-order corrections. 

The right panel of Fig.~\ref{fig:sfbintro} shows the gravitational lensing convergence $\kappa$ and integrated Sachs-Wolfe (ISW) effects, which are quantities projected along the line-of-sight direction. Lensing convergence is the only projected quantity to include angular dependence arising from the Laplacian operator in its definition. 

\begin{figure}
\centering
\includegraphics[width=\textwidth]{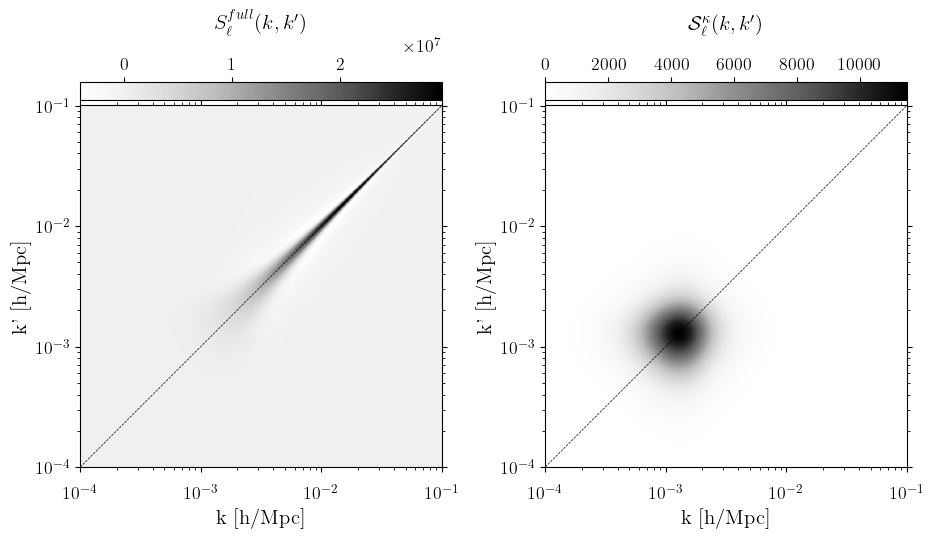}
\caption{\justifying 2D view for the spherical power spectrum. In the left panel the full power spectrum is plotted including all relativistic effects, while the right panel isolates the lensing convergence $\kappa$ contribution. The full power spectrum lays along the diagonal with fast-decaying off-diagonal terms. On the other hand, effects such are lensing are localized in $k$ space and spread oﬀ diagonally.}
\label{fig:vis2d}
\end{figure}

It is also useful to plot the full 2D shape of the spherical power spectrum, as shown in Fig.~\ref{fig:vis2d} for the total one $\mathcal{S}_\ell(k,k')$ in the left panel and the lensing convergence term in the right panel. 
The total power spectrum $\mathcal{S}_\ell(k,k')$ in the left panel is skewed along the diagonal ($k=k'$, i.e.,~same radial distance) while off-diagonal terms ($k\neq k'$) quickly drop to zero. Sec.~\ref{sec:results} explores the relevance of including these off-diagonal modes in the analysis. On the other hand, the right panel shows that effects such as gravitational lensing exhibit a peak around a localized $k$ scale. This exploration of $(k,k')$ space will be further exploited in Sec.~\ref{sec:MW}, to select relevant scales in the multibin cross-correlation spherical power spectrum.

\section{Methodology}
\label{sec:fisher}
We perform our analysis using the Fisher matrix formalism; this is a commonly used tool to forecast performances of a given survey, providing predicted constraints on specific parameters.
In a Bayesian approach, the Fisher matrix can be interpreted as the ensemble average of the log-posterior curvature in its maximum, assuming uniform priors on the parameter~\cite{Tegmark_1997}.
It can also be employed as a tool in experimental design; since it does not depend on specific data realizations, one can compute the matrix for a given likelihood function before performing the actual experiment for a given set of fiducial values for the parameters.

Under the approximation of Gaussianity, the Fisher matrix for $n$ discrete wave vectors $k_i$ can be written as~\cite{beware1}
\begin{equation}\label{sfb_fish_tot}
    F_{\alpha \beta}=\sum_\ell \frac{(2 \ell+1) }{2} \operatorname{Tr}\qty(\hat{\mathbf{C}}_\ell^{-1} \frac{\partial \hat{\mathbf{C}}_\ell}{\partial \theta_\alpha} \hat{\mathbf{C}}_\ell^{-1} \frac{\partial \hat{\mathbf{C}}_\ell}{\partial \theta_\beta})\; .
\end{equation}
The sum is performed over angular modes $\ell$, and the derivatives are performed over cosmological parameters and computed at the fiducial cosmology. The block-diagonal terms are given by

\begin{equation}
\label{eq:defmatcl}
\hat{\mathbf{C}}_{\ell_i}=\qty(\begin{array}{ccc}
\mathcal{S}_{\ell_i}(k_1, k_1) & \cdots & \mathcal{S}_{\ell_i}(k_1, k_n) \\
\vdots & \ddots & \vdots\\
\mathcal{S}_{\ell_i}(k_n, k_1) & \cdots & \mathcal{S}_{\ell_i}(k_n, k_n)
\end{array})\,.
\end{equation}

As usual, $\mathcal{S}_\ell(k_i, k_j)=\Bar{\mathcal{S}}_\ell(k_i, k_j)+\mathcal{N}_\ell(k_i, k_j)$.

Conversely, if the nondiagonal terms are neglected, the covariance matrix is defined in Eq.~\eqref{cov_sfb_kk}, where only equal-$k$ spectra are considered and the Fisher matrix reduces to~\cite{sfb_fish}

\begin{equation}
F_{\alpha \beta}=f_{\text {sky }} \sum_{\ell, i} \frac{(2 \ell+1)}{2} \frac{1}{(\bar{\mathcal{S}}_\ell(k_i, k_i)+\mathcal{N}_\ell(k_i, k_i))^2} \frac{\partial \mathcal{S}_\ell(k_i, k_i)}{\partial \theta_\alpha} \frac{\partial \mathcal{S}_\ell(k_i, k_i)}{\partial \theta_\beta}\,.
\end{equation}
    
Once the FM is computed, the Gaussian covariance matrix of the parameters is retrieved by matrix inversion
\begin{equation}
    \mathrm{Cov}(\theta_\alpha,\theta_\beta)=\qty(F^{-1})_{\alpha\beta}\,,
\end{equation}
and the marginalized uncertainty of a parameter reads
\begin{equation}
    \label{eq:sigmafish}
    \sigma(\theta_\alpha)=\sqrt{\qty(F^{-1})_{\alpha\alpha}}\,.
\end{equation}

\subsection{Covariance}
\label{sec:covoffdiag}
The covariance matrix of the spherical power spectrum is a crucial ingredient to encode the statistical properties of the field. This quantity will be employed to quantify the impact of the relativistic corrections shown in the previous section on the cosmological parameters.

By including terms involving off-diagonal $k$ modes, the covariance matrix embeds correlations between different radial line of sights, providing a more refined description of the statistical features of the field. In the next section, we will show how the inclusion of off-diagonal terms can improve the constraints on cosmological parameters. By neglecting these terms, one loses information on the radial distribution of the field and on correlation due to effects integrated over the line of sight such as lensing convergence, which can be crucial in the analysis of galaxy surveys.

At a fixed multipole $\ell$, $\mathcal{S}_{\ell}(k,k)$ can be estimated by averaging over all $m$ modes and a thin shell $\Delta k$ and subtracting the shot-noise term. Since the $\mathcal{N}_{\ell}$ computation would require precise knowledge of the survey selection function, one can construct the following (unbiased) estimator without performing the noise subtraction~\cite{YD13}:

\begin{equation}
    \hat{\mathcal{S}}_\ell(k, k')=\frac{1}{(2 \ell+1)} \sum_{m} \frac{1}{\Delta k}\frac{1}{\Delta k'} \int_{\Delta k} \int_{\Delta k'}\abs{\delta_{\ell m}(u)\delta_{\ell m}(u')}\,du \, du' \, ,
\end{equation}
where the integration is performed over the band $[k-\Delta k/2,k+\Delta k/2]$. For an extensive analysis of more refined SFB estimators and their properties, refer to~\cite{superfab,sfb_analysis}.

In this section, the spherical Fourier modes $\delta_{\ell m}(k)$ will be assumed to be Gaussian.
Thus, the covariance matrix for $\hat{\mathcal{S}}_{\ell}(k, k')$ reads
\begin{widetext}
    \begin{equation}
        \begin{aligned}\label{covkk_gen}
        &\text{Cov}\qty[\hat{\mathcal{S}}_\ell(k, k'),\hat{\mathcal{S}}_{\ell'}(k'',k''')]=\frac{1}{(2\ell+1)(2\ell'+1)}\sum_{m}\sum_{m'}\frac{1}{(\Delta k)^2}\frac{1}{(\Delta k')^2}  \times \\
        &\int_{\Delta k}\int_{\Delta k'}\int_{\Delta k}\int_{\Delta k'}\Big[\langle\abs{ \delta_{\ell m}(u)\delta_{\ell m}(u') } \abs{ \delta_{\ell' m'}(u'')\delta_{\ell' m'}(u''') }\rangle
        -\langle\abs{  \delta_{\ell m}(u)\delta_{\ell m}(u')  }\rangle\langle\abs{ \delta_{\ell' m'}(u'')\delta_{\ell' m'}(u''') }\rangle\Big]\,du\,du'\,du''\,du'''\,.
    \end{aligned}
    \end{equation}
    \end{widetext}

where the four-point correlator of the spherical Fourier modes is
\begin{equation}
    \langle\abs{\delta_{\ell m}(u)}^2\abs{\delta_{\ell' m'}\qty(u')}^2\rangle=\langle\delta_{\ell m}(k) \delta_{\ell m}^*(k) \delta_{\ell' m'}\qty(k') \delta_{\ell' m'}^*\qty(k')\rangle\,.
\end{equation}
In the assumption of Gaussianity for the overdensity field $\delta(\textbf{x})$, one can rewrite the four-point correlator by making use of Wick's theorem:
\begin{equation}
    \begin{aligned}
    \left\langle\delta\left(\mathbf{x}_1\right) \delta\left(\mathbf{x}_2\right) \delta\left(\mathbf{x}_3\right) \delta\left(\mathbf{x}_4\right)\right\rangle & =\left\langle\delta\left(\mathbf{x}_1\right) \delta\left(\mathbf{x}_2\right)\right\rangle\left\langle\delta\left(\mathbf{x}_3\right) \delta\left(\mathbf{x}_4\right)\right\rangle 
     +\left\langle\delta\left(\mathbf{x}_1\right) \delta\left(\mathbf{x}_4\right)\right\rangle\left\langle\delta\left(\mathbf{x}_2\right) \delta\left(\mathbf{x}_3\right)\right\rangle \\
    & +\left\langle\delta\left(\mathbf{x}_1\right) \delta\left(\mathbf{x}_3\right)\right\rangle\left\langle\delta\left(\mathbf{x}_2\right) \delta\left(\mathbf{x}_4\right)\right\rangle
    \end{aligned}
\end{equation}

In the expansion, note that the $\langle\delta\qty(\mathbf{x}_1) \delta(\mathbf{x}_2)\rangle\langle\delta(\mathbf{x}_3) \delta(\mathbf{x}_4)\rangle$ term is actually equal to the second term in the second line of Eq.~\eqref{covkk_gen},  hence the two disappear in the final expression. The terms $\langle 1,4\rangle$ and $\langle 2,3\rangle$ are constructed as a product of two-point functions computed as in Appendix~\ref{app:2pt}. That is also true for $\langle 1,3\rangle$ and $\langle 2,4\rangle$, with a caveat: one has to be careful to properly convert $Y_{\ell' m'}$, finally retrieving an additional $\delta_{m,-m'}$.

The final expression for the covariance matrix is then

\begin{widetext}
    \begin{equation}\label{cov_sfb_offdiag}
\begin{aligned}
&    \text{Cov}\qty[\hat{\mathcal{S}}_\ell(k, k'),\hat{\mathcal{S}}_{\ell'}(k'',k''')]=\frac{2}{(2\ell+1)}\frac{1}{\Delta k \Delta k'}\frac{1}{\Delta k''\Delta k'''} \\
 &  \times  \int_{\Delta k}\int_{\Delta k'}\int_{\Delta k''}\int_{\Delta k'''} \qty[\mathcal{S}_\ell(u, u''')\mathcal{S}_\ell(u', u'')+
    \mathcal{S}_\ell(u, u'')\mathcal{S}_\ell(u', u''')]\delta_{\ell\ell'}\,du\,du'\,du''\,du'''\,.
\end{aligned}
\end{equation}
\end{widetext}
whereas for compactness $\mathcal{S}_\ell(u, u')\equiv\bar{\mathcal{S}}_\ell(u, u')+\mathcal{N}_\ell(u, u')$. 
For the diagonal terms, in the case of equal and thin $\Delta k$ averaging, the covariance matrix reduces to
\begin{equation}\label{cov_sfb_kk}
\text{Cov}\qty[\hat{\mathcal{S}}_{\ell}(k, k),\hat{\mathcal{S}}_{\ell'}(k',k')] =\frac{2}{2 \ell+1} 
\qty[\bar{\mathcal{S}}_{\ell}(k, k')+\mathcal{N}_{\ell}(k, k')]^2 \delta_{\ell \ell'}\,.
\end{equation}
In the same thin-$\Delta k$ assumption, the full covariance is instead
\begin{equation}\label{cov_sfb_kkprime}
\text{Cov}\qty[\hat{\mathcal{S}}_{\ell}(k, k'),\hat{\mathcal{S}}_{\ell'}(k'',k''')] =\frac{2}{2 \ell+1} 
\qty[\mathcal{S}_{\ell}(k, k''')\mathcal{S}_{\ell}(k', k'')+\mathcal{S}_{\ell}(k, k'')\mathcal{S}_{\ell}(k', k''')] \delta_{\ell \ell'}\,.
\end{equation}
In the following, we refer to the analysis that includes only the diagonal terms of the covariance as the ``diagonal analysis''~\eqref{cov_sfb_kk}, while the approach that also includes off-diagonal terms is labeled as the ``full analysis''~\eqref{cov_sfb_kkprime}. The latter is more computationally expensive, but it provides a more accurate description of the data, as we will show in the next section.

\subsection{Best-fit shift}
\label{sec:BFS}
In this work, not only do we want to provide a full modeling of the two-point SFB spectrum, including all projection effects and the full nondiagonal covariance, but we also quantify what is the impact of using a correct model instead of a simplified approximation (i.e., a Newtonian nonrelativistic description of galaxy clustering). To understand that, we will investigate the change in the signal and the final parameter estimation precision, but also how the inferred best fit of such parameters is affected.

This quantifies the shift in the best-fit parameters (BFS, best-fit shift) due to inaccuracies in the modeling. It is derived from the consideration that any deviation from the true model, whether due to an incorrect assumption of the cosmological model, or due to an imprecise representation of the observable, will result in a bias in the inferred parameters.

The best fit parameters of a model are estimated by maximizing the likelihood of the data given a fiducial model. However, the best fit parameters can be biased if the model is not an accurate representation of the data. This bias can be estimated by expanding the likelihood around the true values of the parameters~\cite{Taylor_2007,Heavens_2007,raccanelli_neutrinos,beware2}. Since the likelihood is not well behaved near the true values of the parameters, the approximation made in the expansion of the likelihood can be inaccurate.

Following the procedure outlined in~\cite{beware2}, here we investigate the impact of including or neglecting GR effects on constraints of cosmological parameters from measurements of the spherical power spectrum $\mathcal{S}_\ell(k,k')$.
The simplified modeling of the data will include only $\delta_m$ and RSD terms [it will be referred to as the RSD term, i.e.,~$\mathcal{S}^{\mathrm{RSD}}_\ell(k,k')$], compared to including all GR effects [labeled as GR i.e.,~$\mathcal{S}^{\mathrm{GR}}_\ell(k,k')$].
In the ``diagonal analysis'', we neglect off-diagonal terms by reducing to $\mathcal{S}_{\ell}(k,k)$.

The shift on a parameter $\theta$ can be written as
\begin{equation}\label{eq:bfs_diag}
    \Delta \theta = \qty(\sum_\ell F_\ell)^{-1}\qty[\sum_{\ell} \frac{\partial \mathcal{S}^{\mathrm{RSD}}_\ell(k,k)}{\partial \theta}\times \operatorname{Cov}^{-1}(k,k')_\ell\times\qty(\mathcal{S}^{\mathrm{GR}}_\ell(k',k')-\mathcal{S}_\ell^{\mathrm{RSD}}(k',k'))]\,,
\end{equation}
where the covariance is computed as in Eq.~\eqref{cov_sfb_kk}. On the other hand, the ``full analysis'' includes off-diagonal terms $\mathcal{S}_\ell(k,k')$. For each angular scale $\ell$ and the relative finite set of radial modes $k_i\in[k_1,\dots,k_{n_\ell}]$, the data can be rearranged in a column vector $\mathcal{C}_{\ell}$, so that
\begin{equation}\label{def:slvec}
\mathcal{C}_{\ell}=\begin{pmatrix}
    \mathcal{S}_{\ell}(k_1,k_1)\\
    \vdots\\
    \mathcal{S}_{\ell}(k_1,k_n)\\
    \vdots\\
    \mathcal{S}_{\ell}(k_n,k_n)
    \end{pmatrix}\,.
\end{equation}
Since $\mathcal{S}_\ell(k_1,k_2)$ is symmetric, only the diagonal and upper-diagonal entries are included. In order to be properly combined with the data vector of Eq.~\eqref{def:slvec}, following~\cite{beware1}, the full 4D covariance is reshaped by associating to every index $I$ and $J$ of the column vector  $\mathcal{C}_{\ell}$ a couple of indices ($I_1$, $I_2$) and ($J_1$, $J_2$), to encode the two $k$ modes ($k_1,k_2$) for each spectrum. The covariance matrix $\mathcal{M}_\ell^{IJ}$ then reads
\begin{equation}
\label{def:Mlijcov}
    \mathcal{M}_\ell^{IJ}=\frac{1}{2 \ell+1}\qty[\mathcal{C}_{\ell}^{\qty(I_1, J_1)} \mathcal{C}_{\ell}^{\qty(I_2, J_2)}+\mathcal{C}_{\ell}^{\qty(I_1, J_2)} \mathcal{C}_{\ell}^{\qty(I_2, J_1)}]\,.
    \end{equation}
One can then write the full expression of the best-fit shift on a parameter $\theta$ as
\begin{equation}\label{eq:bfs_def_comp}
    \Delta \theta = \qty(\sum_\ell F_\ell)^{-1}\qty[\sum_{\ell} \frac{\partial \mathcal{S}^{\mathrm{RSD}}_\ell}{\partial \theta}\times \mathcal{M}^{-1}_{\ell}\times\qty(\mathcal{S}^{\mathrm{GR}}_\ell-\mathcal{S}_\ell^{\mathrm{RSD},J})] \; .
\end{equation}
By construction, the dimensionality of the whole $\mathcal{M}_\ell$ matrix is $N_\ell^2$, with $N_\ell = n_k(\ell)(n_k(\ell)+1)/2$ and $n_k(\ell)$ the number of allowed $k$ modes for each angular scale. This implies a high computational cost in the creation and inversion of the covariance matrix. In order to overcome this, we may select a subsample of relevant $(k,k')$ pairs by isolating a linear strip along the $(k,k)$ diagonal in log-log space and discarding one of the two sides to avoid double counting. The selection is performed for each $\ell$, and the final covariance matrix is obtained by stacking the selected modes for each angular multipole.
In this work, to preserve the full information content of the covariance matrix, we exploit the derivation presented in Appendix A of~\cite{Hamimeche_2008}. In full generality, the expression for the best-fit shift can be remapped to the trace of the product of the matrices in $(k,k')$ space.  This allows us to rewrite the best-fit shift expression in Eq.~\eqref{eq:bfs_def_comp} as 
\begin{equation}
    \label{eq:bfs_def}
    \Delta \theta = \qty(\sum_\ell F_\ell)^{-1}\qty[\sum_{\ell} \frac{2\ell+1}{2}\mathrm{Tr} \qty[ \frac{\partial \mathcal{S}^{\mathrm{RSD}}_\ell(k,k')}{\partial \theta}\times \hat{C}_\ell^{-1}(k,k')\times\qty(\mathcal{S}^{\mathrm{GR}}_\ell(k,k')-\mathcal{S}_\ell^{\mathrm{RSD}}(k,k'))\times \hat{C}_\ell^{-1}(k,k')]]\,,
\end{equation}
where $\hat{C}_\ell(k,k')$ is the covariance matrix in $(k,k')$ space defined in Eq.~\eqref{eq:defmatcl}, and its inversion is now manageable and allows us to retain all the information content of the data without any loss of accuracy.

\subsection{Modes selection}
\label{subsec:modsel}
The theoretical modeling developed in this work makes use of the linear evolution of the density field, as well as a first-order treatment of the GR effects and modifications of a scale-independent galaxy clustering bias. These assumptions are valid only on large linear scales. There are several approaches to compute the maximum linear scale $k_{\rm max}^{3D}(z)$ at each redshift $z$ in the $3D$ Fourier framework. However, the conversion of this threshold to the SFB case is not trivial. For each $k$ mode, an integration is performed over 3D wave numbers $\Tilde{k}$ [see Eq.~\eqref{signal_def}] and including redshift-dependent functions; the $k$, $\ell$, and $z$ dependencies in equations~\eqref{signal_def} and~\eqref{sphmult} make the isolation of clear $k_{\max}(\ell)$ for each $\ell$ not clearly stated. In~\cite{pullen}, each $k_{\max}(\ell)$ is computed by evaluating the fractional difference in the the power spectrum computed using a linear and nonlinear matter power spectrum. For each angular multipole $\ell$, the maximum $k_{\max}(\ell)$ is then retrieved as the scale at which the fractional ratio between the power spectrum computed using a linear and nonlinear prescription is less then a threshold value set at $10\%$. The work in~\cite{pullen} shows that for the values of $\ell$ included in this work ($\ell\lesssim 150$), the $\ell$ dependence of $k_{\max}(\ell)$ is negligible, and it will hence be fixed at $k_{\max}=0.15\,\mathrm{h/Mpc}$. Concerning the  $\Tilde{k}$ integration in Eq.~\eqref{signal_def}, linear theory will be assumed for the whole integration range.

On the other hand, the minimum $k_{\min}(\ell)$ for each $\ell$ is determined by the radial coverage of the survey. For a finite survey with a sharp cutoff $r_{\rm max}=R$, this additional boundary condition implies a discrete spectrum of modes $k=k_{n,\ell}$, which arises for each angular mode $\ell$ to keep the SFB basis function orthogonal~\cite{fish95}. As discussed in Appendix~\ref{app:kdisc}, this allows one to select a finite number $n_{k}(\ell)$ of $k$ modes to include in the analysis, in the range $k\in [k_{\min }(\ell),k_{\max}(\ell)]$.  Refer to~\cite{Heavens_1995,fish95} and Appendix~\ref{app:kdisc} for further details.

In this section, we do not analyze top-hat-bounded surveys. In order to still select physically meaningful $k$ modes, we generalize $R$ in terms of the comoving survey volume $V(r)$ as
\begin{equation}\label{defrmax}
    V(R) = \int_0^R\int_0^{\pi}\int_0^{2\pi} \phi(r)r^2\sin(\theta)\,d\phi\,d\theta\,dr=0.90V_{\infty}\,,
\end{equation}
with $V_{\infty}$ the full volume integral of the survey selection function $\phi(r)$.


\subsection{Survey specifications}\label{surveyspec}

\begin{figure}
    \centering
\includegraphics[width=\textwidth]{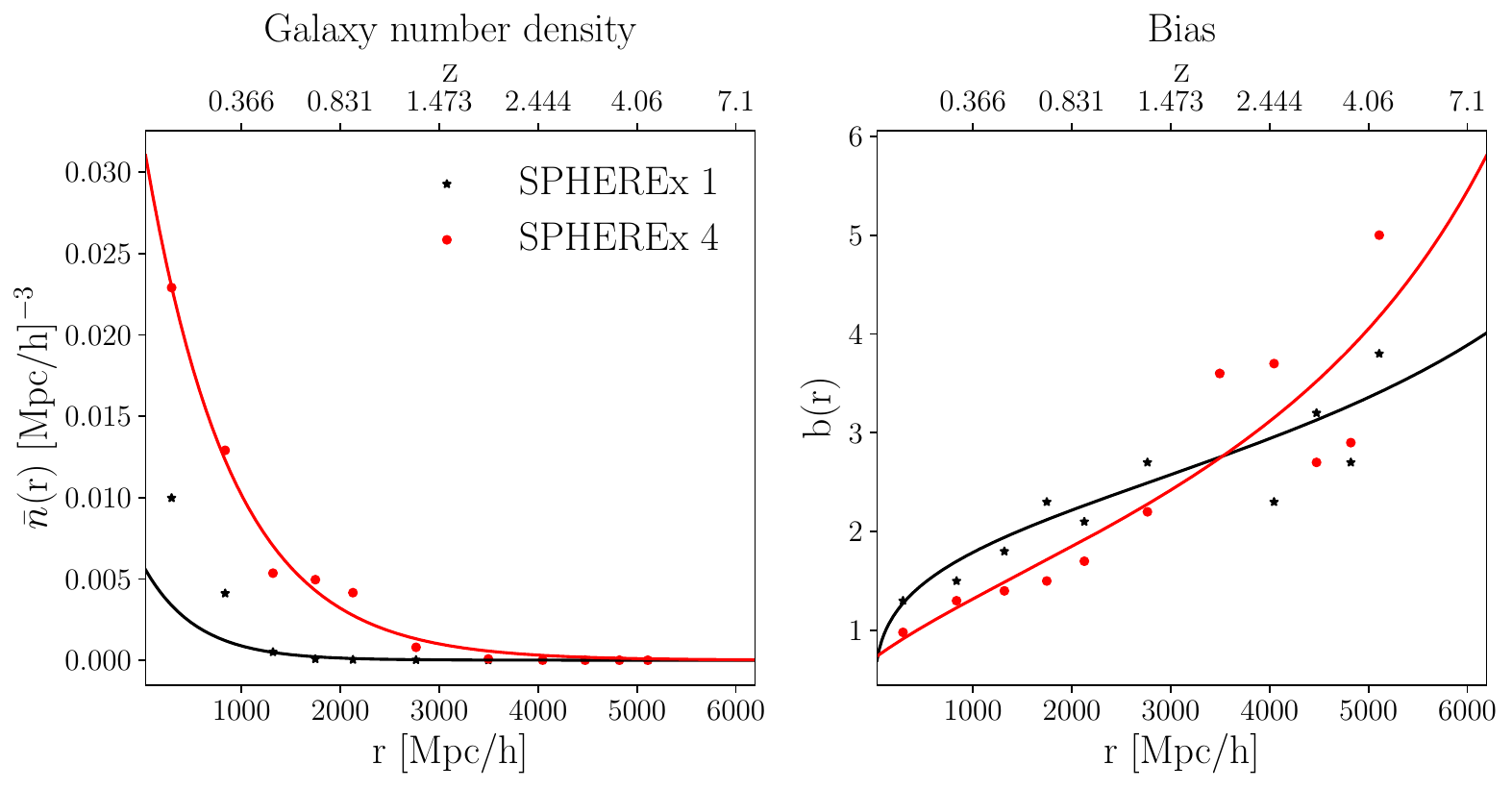}
    \caption{\justifying Galaxy number density (left panel) and bias (right panel) used in the analysis, corresponding to different redshift accuracy in the redshift measurement for a SPHEREx-like survey. Fitting templates have been used to build continuous functions from the data points available for a discrete set of redshift bins. }
    \label{fig:spherex_dist}
\end{figure}
As a reference survey for our study, we use the planned NASA space mission SPHEREx~\cite{spherex_cosmo}. This upcoming survey will provide a large number of sources at low redshift and over the full sky, making it a proper test bench for large-scale GR projection effects. We model it using public data from~\cite{spherex_cosmo} as reference galaxy radial distribution and bias, as shown in Fig.~\ref{fig:spherex_dist}. If not otherwise stated, the analysis will be performed for the SPHEREx4 distribution. The results do not represent an actual SPHEREx forecast and we generalize the analyses to hold for a generic full sky. More details of the distribution uncertainties are discussed in Sec.~\ref{sec:bias}.

The public data provide values for each redshift bin, which we fit using
\begin{equation}
    \bar{n}(r,a,b,c)=a\exp\qty[-\qty(\frac{r}{b})^c]\,,
\end{equation}
where $r$ is the comoving distance.

Similarly, the scale-independent bias has been fitted as
\begin{equation}
\label{bias_fit}    b(r,a_{\mathrm{bias}},b_{\mathrm{bias}},c_{\mathrm{bias}})=a_{\mathrm{bias}}[1+b_{\mathrm{bias}}\cdot z(r)]^{c_{\mathrm{bias}}}\,,
\end{equation} 
where $z(r)$ is effectively computed as an interpolation function that connects a given comoving distance $r$ to its associated redshift (note that this is cosmology dependent, and the function is recomputed when varying cosmological parameters). For the SPHEREx4 sample, $[a_{\mathrm{bias}},b_{\mathrm{bias}},c_{\mathrm{bias}}]=[0.72, 0.18, 0.55]$. By marginalizing over the three fitting parameters one can then include the uncertainty of the bias modeling in the analysis, allowing for flexibility of the bias template function. 

In realistic surveys, when building the measured galaxy number density distribution $n(z)$, in addition to instrument-dependent features, one has to also account for redshift evolutions. Additional contributions arise since new galaxies form (i.e.,~the true number density of galaxies does not scale as the inverse cube of the cosmological scale factor $a^{-3}$) and the galaxy sample is flux limited. The luminosity
threshold $L_{\mathrm{min}}$ for a detection is connected to the observed flux as in equation \eqref{app:lmin}. As discussed in Appendix~\ref{app:magbias}, the observed mean number density of galaxies then reads $\bar{n}_\text{g}(z)=n(L_\text{lim}(z),z)$, and we can define the magnification bias as 
\begin{equation}
    Q(z)=-\left.\frac{\partial \ln \bar{n}\left(L_{\min }, z\right)}{\partial \ln L_{\min }}\right|_{L_{\min }=L_{\lim }(z)}\,.
\end{equation}
We will present the results in terms of $s=2Q/5$, which arises when recasting the magnification bias from luminosity to the observed magnitude (more details in Appendix~\ref{app:magbias}). Due to the lack of accurate data, for this preliminary study we model the magnification bias as a redshift-independent quantity $s=0.6$, as suggested in~\cite{beware2,Bosi_2023}. The robustness of this approximation will be tested by varying the magnification bias over a range of both negative and positive values.

\section{Impact of GR effects on parameters estimation}
\label{sec:results}
Now we turn to evaluating the magnitude and relevance of relativistic corrections to the observed spherical power spectrum.
Following the previous section, failing to correctly model the total signal by not including projection effects may result in a misestimation of cosmological parameters, affecting both the best fits and the predicted error bars.
Our focus here will be on how relativistic effects may impact the constraints, for a given survey, on PNG, as its effect is degenerate with some of the effects considered. 

The impact of non-Gaussian statistics on the large-scale structure of the Universe has been widely studied in the literature, e.g.~\cite{sabino_png,salopek91,verde_sabino}. Observational constraints from CMB have been discussed in, e.g.,~\cite{gangui94,Bartolo_2004,planckcollaboration2019planck},
and various analyses of PNG measurements in the context of radio and spectroscopic galaxy surveys have been performed in e.g.~\cite{Raccanelli_2015,RACCANELLI201735,Mueller_2018}.

Deviations from Gaussianity can be parametrized as $\Phi_{\mathrm{NG}}=\phi+\fnl(\phi^2-\langle\phi^2\rangle)$, where $\phi$ is a Gaussian random field and $\fnl$ encodes the amplitude of the quadratic correction of the potential~\cite{Komatsu_2001}. 
A nonzero $\fnl$ induces a scale-dependent term in the large-scale halo bias~\cite{Dalal_2008,Matarrese_2008}. This has been studied in relation to the general-relativistic description of galaxy clustering in, e.g.,~\cite{Desjacques_2010,Jeong_2012,Bruni_2012,Raccanelli_2013,Raccanelli_2016}. The full, non-Gaussian bias then reads 
\begin{equation}
    b_{\mathrm{NG}}(k,z)=b_G(z)+2(b_G(z)-p)\fnl \frac{3\Omega_m}{2D\left(z\right)}\frac{g(0)}{g(z_{\mathrm{rad}})}\frac{H_0^2}{k^2T(k)}\delta_c\,,
\end{equation}
where $b_G(z)$ is the usual Gaussian bias parameter and $D(z)$ is the linear growth factor normalized at $z=0$. This choice of normalization requires to introduce the additional ratio $g(0)/g(z_{\rm rad})\simeq 1.3$ with $g(z)=(1+z)D(z)$~\cite{Riquelme_2023}. The factor $\delta_{\mathrm{c}}=1.686$ is the critical overdensity for spherical gravitational collapse in a Einstein–de Sitter cosmology~\cite{Sheth_1999,Sheth_2001}.
There has been considerable attention on the value of $p$ in recent years, and how accurate it is to set $p=1$~\cite{png}. The topic is still under debate, with some possible ways to mitigate its impact suggested in~\cite{liciaP}.
It is beyond the scope of this paper to delve into this in detail, therefore we fix here $p=1$, and it will be straightforward to adapt our conclusions to the case where some uncertainty in $p$ in included.

Both the linear growth factor $D(z)$ and the linear transfer function $T(k)$ are normalized to $1$ at $z=0$ and $k=0$ and the bias is defined with respect to the comoving gauge matter density. It should be noted that, as pointed out in~\cite{Bahr_Kalus_2021}, the peculiar motion of the observer, if not accurately accounted for, significantly biases the inference of PNG parameters. Furthermore, GR effects such as lensing convergence are degenerate with $\fnl$ in terms of their $k$ dependence in the spherical power spectrum~\cite{raccanelli2016doppler}.

This implies that by neglecting GR effects, the inferred $\fnl$ value may be systematically biased. In the next sections, we aim to quantify this shift and its relevance compared to the actual precision of the inference.
For our calculations, we set the fiducial value for local non-Gaussianity to $\fnl^{\rm fid} = 0.1$~\cite{planckcollaboration2019planck}.

\subsection{Impact on $\fnl$}
We can now perform our investigation for a SPHEREx-like survey, choosing the two subsamples presented in Fig.~\ref{fig:spherex_dist}. SPHEREx 4 represents the subsample with the highest galaxy number count: an overall higher number density allows us to reduce the shot noise and hence improving the precision of the statistical constraints over cosmological parameters. We also test our results on the SPHEREx 1 subsample with a lower number density of galaxies, reduced radial depth and higher redshift precision. We do not model redshift accuracy in the spherical power spectrum. The aim is to explore the BFS on $\fnl$ across different number density functional shapes. It should be pointed out that these do not represent actual SPHEREx forecasts. 

\begin{figure}
\centering    \includegraphics[width=0.75\textwidth]{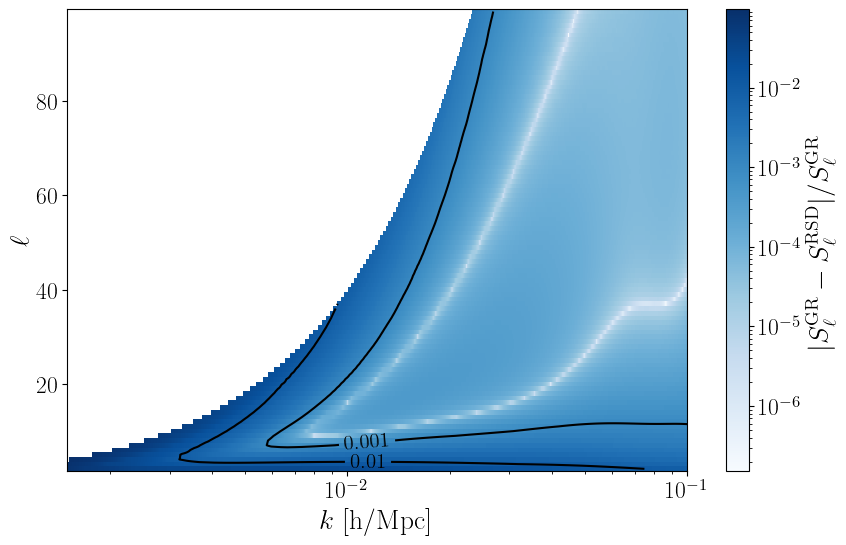}
\caption{\justifying Relative difference of the full spherical power spectrum including all effects (GR) with respect to the one including only density and RSD. Black contour lines limit the $1\%$ and $0.1\%$ region, respectively. For smaller angular scales, GR effects do not affect considerably the power spectrum, hence the two modelings are increasingly difficult to differentiate. 
The relative difference turns negative inside the white contour, and the blank region is due to the boundary conditions discussed in the text.  
}
\label{fig:compare_rsd}
\end{figure}

We start by comparing the full GR spherical power spectrum to the Newtonian one. We employ the notation introduced in Sec.~\ref{sec:BFS} to label the spherical power spectrum computed accounting for all projection and relativistic effects (GR) and the Newtonian one (RSD) which includes just the terms from matter overdensity and RSD. In Fig.~\ref{fig:compare_rsd} we plot the relative difference between the GR and RSD spherical power spectra as a function of angular modes $\ell$ and radial modes $k$. The blank region is a consequence of the $k_{\min}^{\ell}$ set by the boundary conditions as discussed in Appendix~\ref{app:kdisc}.
One can immediately see that the relative difference is larger for larger angular scales, as expected. For smaller angular scales, GR effects do not considerably affect the power spectrum, and the simplified modeling holds. Indeed, higher $\ell$s show a relative difference of the order of $10^{-5}$ for large $k$. However, in Eq.~\eqref{eq:bfs_def}, a multiplicative factor $(2\ell +1)/2$ enters the sum over angular modes. Including this multiplicative factor to the relative difference, the is still of the order $10^{-3}$ for large $\ell$, leading to smaller contributions to the total best-fit shift.  

\subsubsection{Full analysis}
The main aim of this section is to quantify the impact of neglecting GR effects on the inference of cosmological parameters, focusing on the PNG parameter $\fnl$, which enters the overdensity field and the spherical power spectrum with the same $k$ dependence as some of the GR effects considered, leading to potentially relevant degeneracies. We make use of the formalism introduced in Sec.~\ref{sec:BFS}: the BFS is then compared to the predicted precision for the same parameter (computed through a Fisher-matrix approach) to quantify the relevance to the shift. 

\begin{figure}
\centerline{
\includegraphics[width=\textwidth]{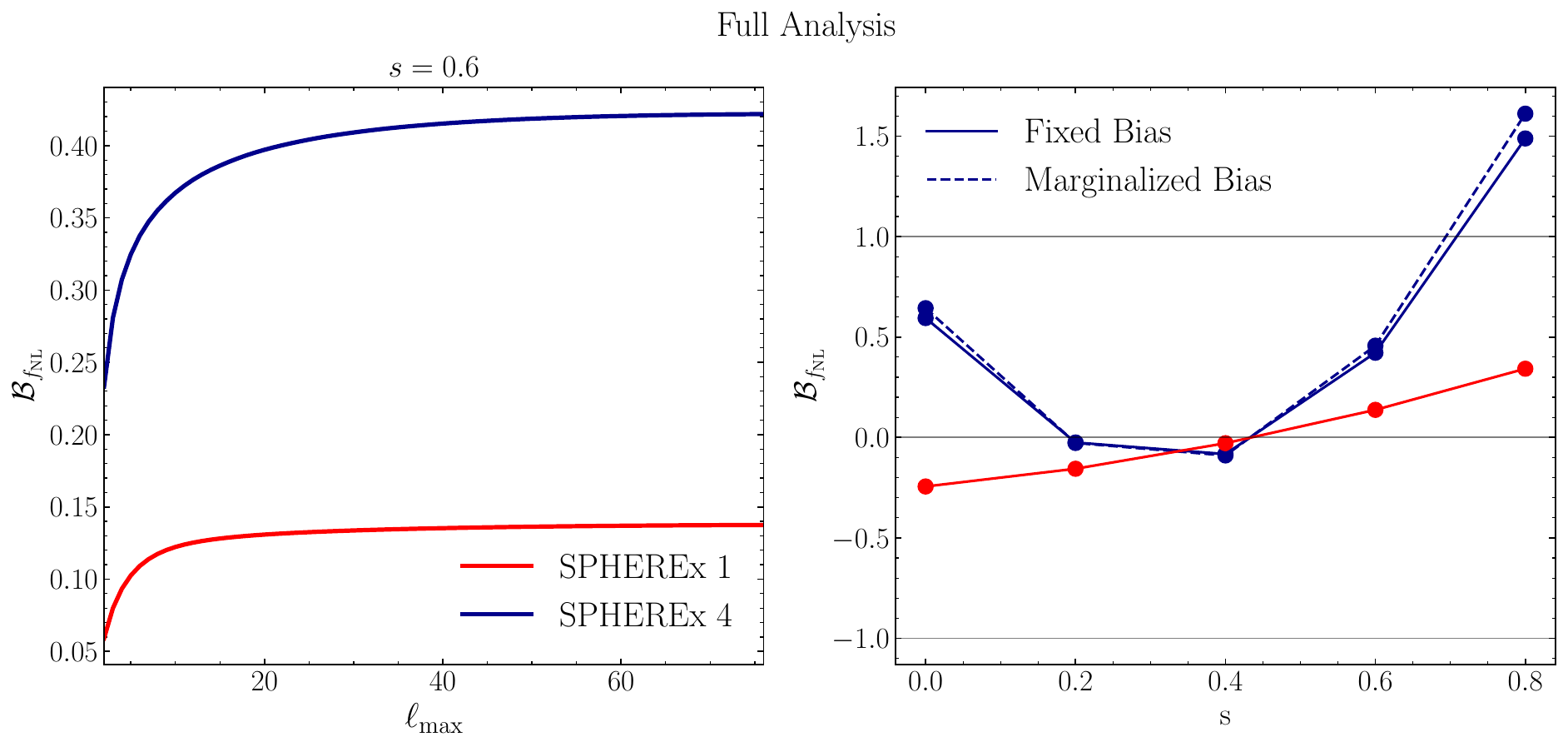}}
\caption{\justifying Left panel: ratio $\mathcal{B}_{\fnl}\equiv\Delta \fnl / \sigma_{\fnl}$ between the best-fit shift $\Delta\fnl$ introduced by neglecting GR effects and the predicted precision on measuring the local non-Gaussianity parameter $\fnl$. This is shown for the two SPHEREx-like subsamples in Fig.~\ref{fig:spherex_dist} and fixed magnification bias $s=0.6$.  The shift asymptotically saturates to a constant value as higher angular modes are less sensitive to GR effects (see also Fig.~\ref{fig:compare_rsd}). Right panel: the ratio $\mathcal{B}_{\fnl}$ for a set of constant magnification bias values $s$. Setting $s=0.4$ totally removes any lensing contribution (the most relevant effects), hence GR effects are highly suppressed and GR and not-GR spherical power spectra show extremely similar features. Bias marginalization is shown for the SPHEREx4 sample. The magnitude of $\mathcal{B}_{\fnl}$ is survey dependent and the figure shows that a BFS is systematically present in the analysis.}
\label{fig:bfs2d}
\end{figure}

The left panel of Fig.~\ref{fig:bfs2d} shows the ratio $\mathcal{B}_{\fnl}\equiv\Delta \fnl / \sigma_{\fnl}$ for increasing maximum angular multipole $\ell_{\max}$ [i.e.,~the maximum $\ell$ included in the summation both in Eqs.~\eqref{sfb_fish_tot} and~\eqref{eq:bfs_def}], between the best-fit shift $\Delta \fnl$ introduced on $\fnl$ by neglecting GR effects  (using instead only $\delta_{ m }$ and RSD terms) and the $68\%$ confidence level $\sigma_{\fnl}$ on the same parameter. $\Delta \fnl$ encodes the accuracy of the simplistic modeling that neglects GR effects, which is compared to the precision of the parameter constraint $\sigma_{\fnl}$. The ratio is shown for both samples considered, for a fixed set of cosmological parameters (fixed at the Planck values~\cite{planckcollaboration2019planck}) and varying $\fnl$. The spherical power spectrum $\mathcal{S}_\ell(k,k')$ includes both diagonal and off-diagonal terms. This choice has been made in order to focus our attention on the effects relativistic corrections have on the estimates of PNG. Given that relativistic corrections are largest at the largest scales, we expect little effects on other parameters. Moreover, while a full evaluation will be appropriate for real data analyses, those results will be heavily dependent on particular aspects of the dataset to be analyzed, and therefore it is not part of the scope of this work. 

As it can be seen in the left panel of Fig.~\ref{fig:bfs2d}, the ratios $\mathcal{B}_{\fnl}$ computed for the two subsamples show similar trends, asymptotically saturating to a constant value. This is due to the fact that the shift is driven by large angular scales, which are more sensitive to GR effects. As more angular modes are included in the analysis, the precision of the constraint increases. On the other hand, the added angular modes are less informative (i.e.,~the difference term inside round brackets in Eq.~\eqref{eq:bfs_def} approaches zero, as discussed in Fig.~\ref{fig:compare_rsd}), hence both the BFS and the $1\sigma$ confidence level saturate to a constant value. The ratio for the high-number-count sample saturates to around $\mathcal{B}_{\fnl}\sim 0.4$, showing that the BFS is relevant in this context. A lower number of sources reduces the relevance of the BFS to around $\mathcal{B}_{\fnl}\sim 0.14$, but a systematic shift is still present. 

In order to test the robustness of our results when including uncertainties in the actual observed distribution, in the following sections, we analyze how the BFS is affected when changing the biasing schemes $b(k,z)$, $s(z)$.


\subsubsection{Biases and distribution uncertainties}
\label{sec:bias}

We fit the galaxy bias from the public SPHEREx information, using a set of parameters as in Eq.~\eqref{bias_fit}. In this work, we marginalize over the three bias parameters and address their impact on the constraints on $\fnl$ and on the total BFS. Up to this point, the BFS has been computed assuming perfect knowledge of the bias model, while the marginalization process allows us to include the uncertainty over bias modeling in the analysis.
Since the bias is degenerate with the $\fnl$ parameter, this will result in a degradation of the constraints on $\fnl$ with respect to a fixed-bias analysis.
While we assume a functional shape for the bias, we also want to remain as general as possible. In a standard clustering analysis with redshift bins, one can leave the bias free in each bin. In our case, the way to remain agnostic about the bias is to marginalize over all possible values of $a_{\mathrm{bias}},b_{\mathrm{bias}},c_{\mathrm{bias}}$. For more details, see Appendix~\ref{app:contour}, where we show the full contour plots for the cases analyzed and discuss the impact of the bias parameters on the BFS. 2D contour plots are also useful to better visualize the BFS when not projected to a one-dimensional submanifold of the global parameter space. The full BFS should be computed in a $n_{\rm par}$-dimensional space and the confidence ellipses allow us to explore how the BFS might vary along directions orthogonal to parameter degeneracies.

We also quantify the impact of uncertainties in the modeling of magnification bias. As discussed in Sec.~\ref{surveyspec}, we assume a constant $s=0.6$ (see~\cite{Bosi_2023}). We verify the importance of this assumption by varying the magnification bias over a range of values, instead of performing a full marginalization over $s$. As discussed in Appendix~\ref{app:GR} (specifically refer to Eq.~\eqref{eq:GR_terms}), the magnification bias enters the lensing convergence term $\kappa$ and is degenerate with $\fnl$ in the spherical power spectrum. This implies that the BFS will be affected by the choice of $s$.
The right panel of Fig.~\ref{fig:bfs2d} shows results of the analysis for $\ell_{ \max }= 80$, given that we saw above the BFS is by then saturated. The magnification bias enters the lensing convergence term $\kappa$ as $\delta_\kappa\propto(5s(z)-2)\kappa$, hence for $s=0.4$ lensing suppression implies close-to-zero best-fit shifts, since among GR effects that may create degeneracies with $\fnl$, the lensing contribution is the most relevant. This is reflected on the $\mathcal{B}_{\fnl}\sim 0$ for both samples.
The BFS is less relevant for the shallower survey, which is less sensitive to GR integrated effects, as incoming photons travel shorter distances and are less affected by the lensing convergence. 

The behavior of the BFS ratio differs between survey configurations due to a complex interplay of all kernels in the spherical power spectrum. The full signal is composed of all combinations of all kernels $\mathcal{M}^i_\ell(\tilde{k},k)$ (as described in Section 3), which are weighted by the number density $n(z)$, the bias $b(z,k)$, and the evolution bias $b_e(z)$. The weighting of these kernels affects the relative relevance of single combinations, and in turn the behavior of the BFS ratio.

However, a systematic shift is present for all configurations analyzed. As discussed in Sec.~\ref{sec:BFS}, the connection between survey propeties and BFS is not straightforward: different number densities imply different shot noise levels, which in turn affect the BFS both in the term proportional to $F^{-1}$ and in the covariance matrix [compare Eq.~\eqref{eq:bfs_def} and~\eqref{sfb_fish_tot}]. Furthermore, bias and magnification bias introduce additional subtle dependencies on the overdensity field. This implies that the numerical results of our BFS analysis should not be interpreted in a quantitative perspective, as the actual BFS will depend on the specific survey specifications and galaxy population properties. Nonetheless, the presence of a systematic BFS is a fully general result that provides a cautionary tale for future inference analyses. We focused on the impact of GR effects but the same reasoning holds for any other potential source of systematics at the observation or theory level, whose joint contribution may affect the final cosmological parameter estimation. 

In Appendix~\ref{app:diagonal}, we show the same analysis performed using only the diagonal terms of the spherical power spectrum. The results are consistent with the full analysis, showing that the BFS is consistently present when varying both the biasing scheme and the galaxy population properties.
However, one should be careful when comparing the numerical results of the two analyses. The single BFS estimates the systematic error induced by using two theoretical modelings that connect the same summary statistics of our observable (the spherical power spectrum $\mathcal{S}_\ell$) to a set of parameters through a different set of assumptions and approximations. This is self-consistent within the same analysis, but the comparison between full and diagonal analysis may be misleading, as that would imply comparing two inferred parameters from two different statistical analyses (both encoding $\mathcal{S}_\ell$) and not referring them to the same underlying physical quantity.

\section{MultiWindow approach}
\label{sec:MW}
Up to now, we considered correlations for galaxy pairs over
the whole survey.
However, it might be convenient (or even necessary) to separate the observed catalog into several radial bins. In that case, it will be necessary to consider correlations that are both in the same bin (auto-correlations) and across different redshift bins (cross-bin correlations).

As an example, we consider a two-bins configuration, where the bins are defined by two top-hat window functions $W_i(z)\in[z_1,z_2]$.
This however introduces the presence of oscillations on the spherical power spectrum, due to the finite integration boundary for the spherical multipoles $\mathcal{M}_\ell$, as discussed in the next sections and visualized in Appendix~\ref{app:MW} (specifically, see Fig.~\ref{fig:mwsurv}). To partially overcome this issue, we model the top-hat window functions as double hyperbolic-tangent functions, with variable parameters to define the width and position of the bins as well as the tilt of their boundaries. This allows us to have a smooth transition that mitigates oscillations. On the other hand, the tail of the window functions is not as sharp as the top-hat one, and one must be careful when interpreting the results for separated bins.
Each spherical multipole defined in Eq.~\eqref{sph_mult_win}, will now be computed for both window functions. Latin letters now label the redshift bin index. The spherical power spectrum is then defined by four indexes $(k,k',i,j)$ and reads
\begin{equation}
\label{MWpowsp}
\mathcal{S}_\ell^{ij}(k, k')=4 \pi \int d \ln \tilde{k} \Delta(\tilde{k}) \mathcal{M}_\ell^{i}(\tilde{k}, k) \mathcal{M}_\ell^{j}(\tilde{k}, k') \,.
\end{equation}

By considering only the $\delta_{ m }$ and RSD terms, all contributions are intrinsically local and the cross-bin power spectrum would be suppressed above a certain radial coherence length.
On the other hand,  
integrated effects such as lensing convergence $\kappa$ introduce nonlocal contributions, hence a nonvanishing cross-bin power spectrum~\cite{Raccanelli_2016}. This means that the multiwindow approach with cross-bin correlations could be employed to isolate and extract projection effects from the total signal.

We test this for a set of log-spaced $(k,k')$ modes to showcase the theoretical framework. To implement this in an actual analysis, one should first select the proper $k$ modes as discussed in Sec.~\ref{subsec:modsel}, both to preserve linear scales and account for the power spectrum discretization due to a finite spherical shell. 

\begin{figure}
    \centering
    \includegraphics[width=\linewidth]{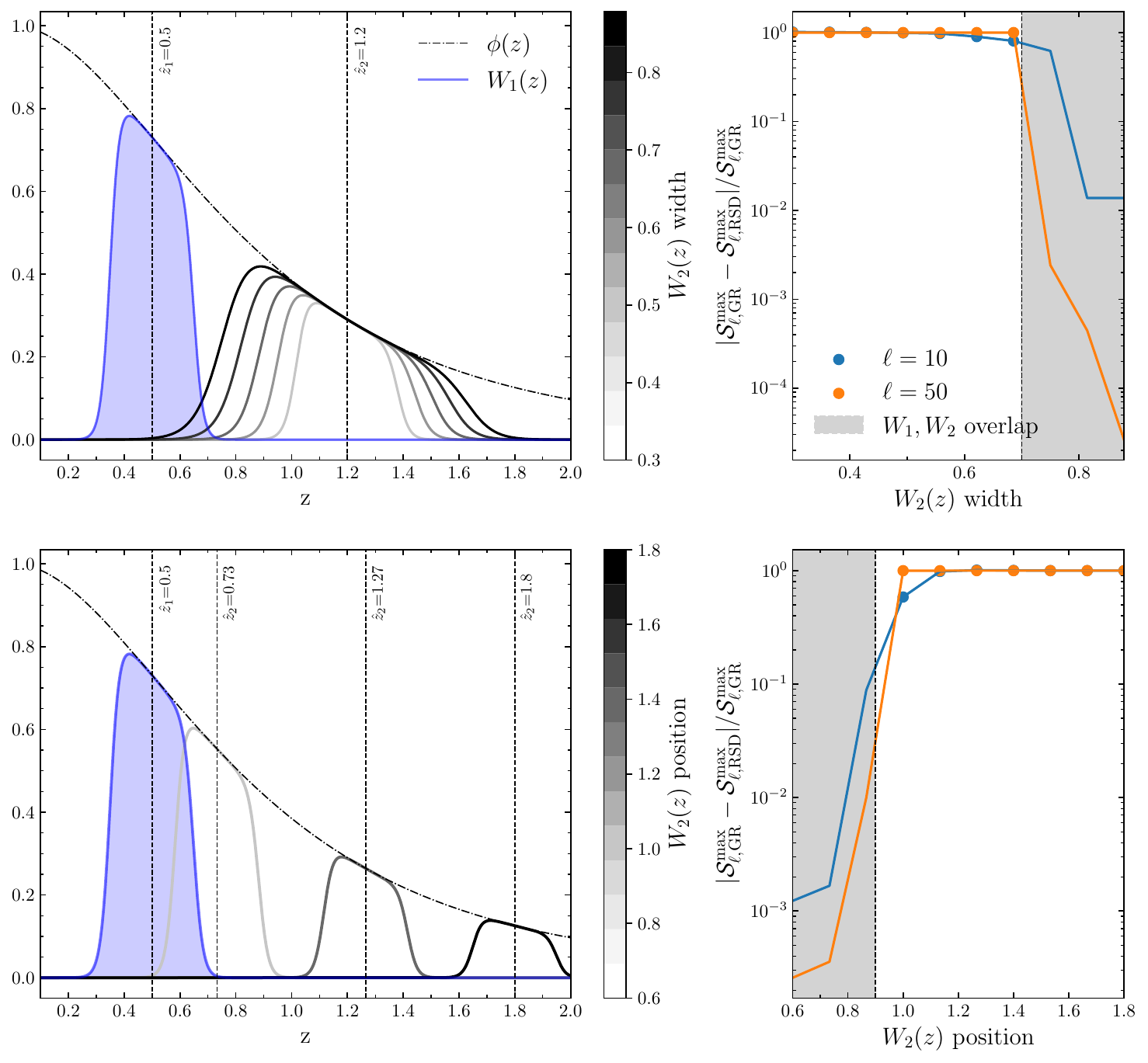}
    \caption{\justifying Multiwindow approach. Left panels: reference Gaussian selection function $\phi(r)$ and two bins. In both cases, the bin closer to the observer is fixed. In the top row, the second bin spans a range of widths, going from well separated to overlapping. In the bottom row, the second bin position is varied from a partly overlapping configuration to a well-separated one. Right panels: relative difference of the peak value of the cross bin power spectrum for $\ell=10$, comparing the full GR case and the Newtonian RSD one. The fully GR power spectrum dominates when the two bins are separated due to the presence of integrated terms. The detection of such signal would be a clear smoking gun of GR effects.}
    \label{fig:mw1}
\end{figure}

Fig.~\ref{fig:mw1} provides a visual representation of this process. The left panels show a reference Gaussian selection function $\phi(z)$ (with shape and features defined in Eq.~\eqref{eq:gaussprova}) and two (almost) top-hat bins.
In the top row, the bin closer to the observer (in blue, where $\hat{z}_1$ is the center of the first bin) is fixed, while the other spans a range of widths, keeping the center fixed at $\hat{z}_2$. Different shades of gray indicate different widths, from light to dark.
In the bottom row, the position of the second bin varies while fixing their widths. Here $\hat{z}_1$ still indicates the center of the first bin, while shades of grey indicate the position of the center of the second bin, $\hat{z}_2^i$.
In both cases, the configurations move from a well-separated bin scenario to a (non total-) overlapping one.
The right panels show the relative difference of the peak value of the cross-bin power spectrum for $\ell=10,50$, comparing the full GR case with the Newtonian RSD one. We choose these reference $\ell$s to explore the main feature of the signal for different angular scales. In scenarios with substantial bin overlap, local terms dominate the power spectrum signal, with density correlations and RSD effects being the primary contributors. This leads to GR effects becoming less significant, resulting in smaller relative differences between the full GR and Newtonian RSD power spectra. Conversely, for well-separated bins, the integrated terms along the line of sight become the dominant contribution, allowing GR effects to drive the signal and maximize the relative differences.

An expansion of this formalism including an extensive detectability analysis is left for future work. The aim here is to explore how isolate clear imprints of GR effects in the clustering signal.

Finite differences computed for higher $\ell$ modes show a similar behavior, but the spectra themselves are less relevant since the signal is suppressed by the finite window depth. 
On the other hand, when the two bins overlap, the total signal almost matches the Newtonian RSD case, as the $\delta_{ m }$ and RSD terms are the dominant contributions and the GR effects are subdominant, as discussed in the previous section.

In Fig.~\ref{fig:mwtanh} in Appendix~\ref{app:MW}, we show the full power spectrum for the same configurations, to provide a complete depiction of its structure at fixed angular scale. In the following sections, we provide an intuitive visualization of this cross-bin power spectrum, highlighting how it allows us to target specific regions of $(k,k')$ space.


\subsection{Limber approximation}
\label{sec:limber}
The full calculation of the spherical power spectrum can be computationally very expensive. In order to try to tackle this issue, the Limber approximation~\cite{limber,kaiser92,LoVerde_2008} is often used. To implement this approximation, one assumes small angular separations (i.e.,~large $\ell$) and that the integrand is composed of highly oscillating functions combined to slower varying terms. In this framework, the spherical Bessel function $j_\ell(x)$ is approximated as a Dirac delta function peaking at $x\approx\nu\equiv \ell+1/2$ (for some discussions on this, see also~\cite{2fast, raccanelli2023power}) and normalized as
    \begin{equation}
        j_\ell(kr)\sim \sqrt{\frac{\pi}{2\nu}}\delta^D(\nu-kr)\,.
    \end{equation}
Our results are compared with those using the Limber approximation (we show in Appendix~\ref{app:MW}, the derivation in this case), to investigate whether it is a viable option for this type of analysis.

\begin{figure}
    \centering
    \includegraphics[width=\linewidth]{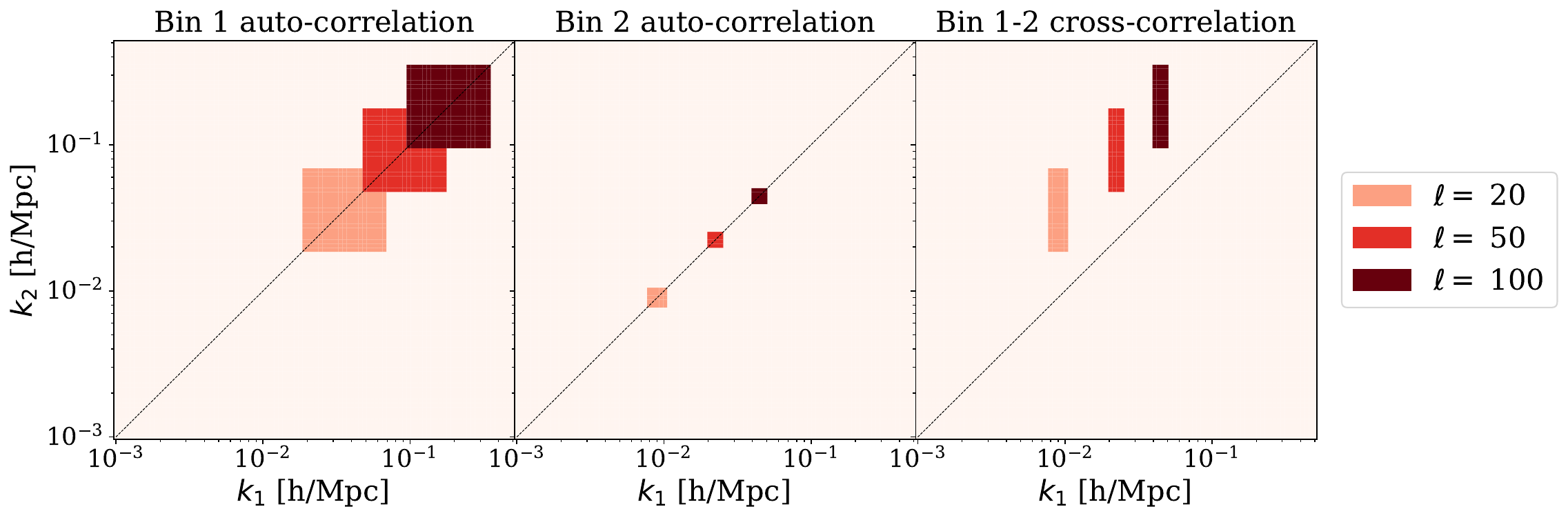}
    \caption{\justifying 2D visualization of the correlation region in $k$ space i.e.,~where the correlation is nonvanishing in the Limber approximation for different $\ell$ modes. The autocorrelation power spectrum is localized along the diagonal while the cross-correlation one is located purely off-diagonally. The full non-Limber computation still preserves correlations outside the colored regions, which are used to provide intuition and inform the choice of $k$ modes to be numerically evaluated}
    \label{fig:mwvis}
\end{figure}

In the Limber approximation, the integrations along the line of sight are suppressed by the presence of the Dirac delta function. Thus, the spherical multiwindow power spectrum in this approximation is formally nonvanishing in a specific area of $k$ space given by the product of the two window functions. This is visualized in Fig.~\ref{fig:mwvis}: the autocorrelation spectra are localized along the diagonal, while cross-correlation ones are pure off-diagonal quantities.
It should be stressed that the full non-Limber computation preserves correlations outside the colored regions, although of lower amplitude. The colored regions provide intuition and inform the choice of $k$ modes to be numerically evaluated.

\subsubsection{Integrated terms and local-integrated mixed terms}

\begin{figure}
    \centering
    \includegraphics[width=\linewidth]{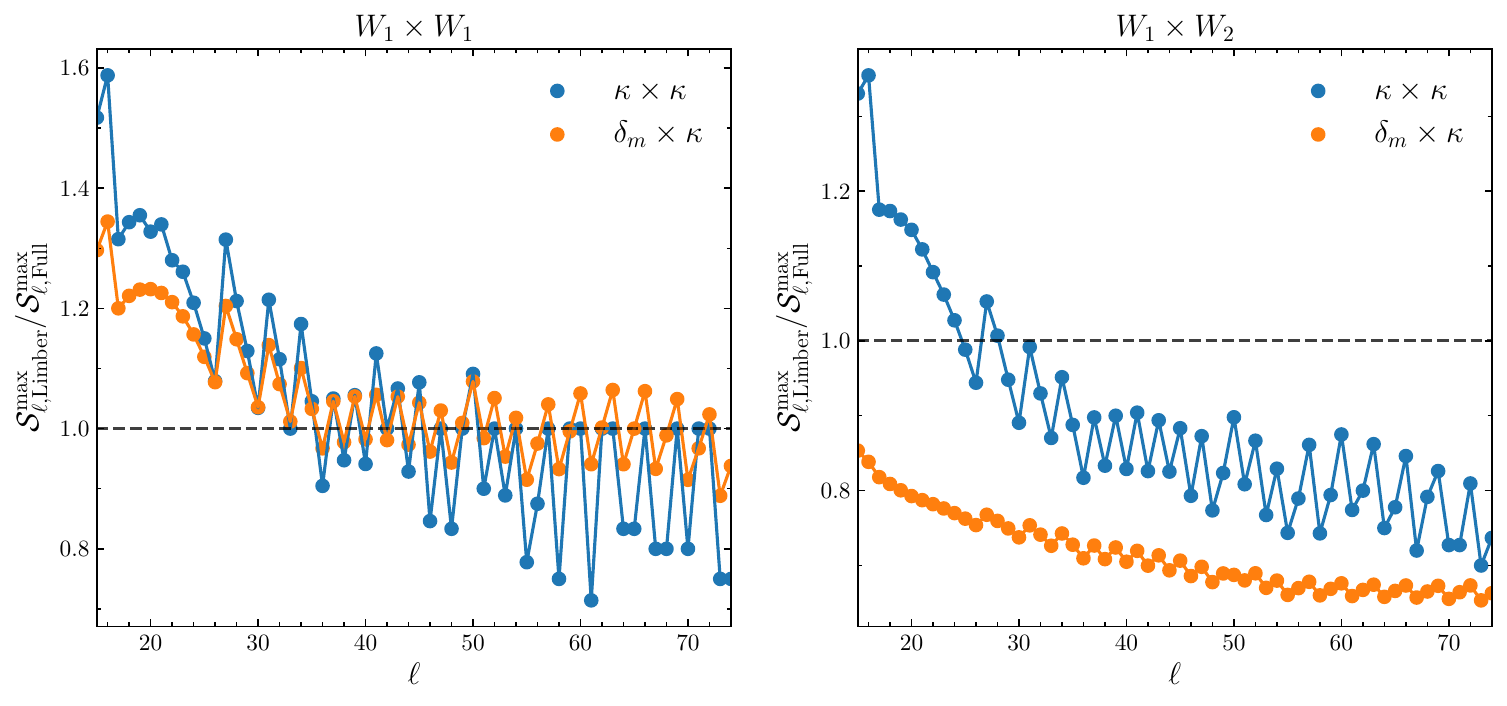}
    \caption{\justifying Ratio between the full computation and the Limber approximation for increasing $\ell$ and two combinations of local and integrated effects. The left panel shows the autocorrelation within the same bin, while the right panel shows the cross-correlation between separated redshift bins. The oscillations are due to the discrete integration over the spherical Bessel functions and the discrete set of $k$ modes over which the integration is performed. Limber approximation breaks down for lower $\ell$s as the spherical Bessel function is not well approximated by a Dirac delta function. The cross-bin spectrum is instead affected by the Limber approximation and the ratio is consistently lower than the same-bin correlation. The full computation is crucial in this case, as it preserves terms integrated along the line of sight, which are not well modeled by the Limber approximation.}
    \label{fig:limbratio}
    \end{figure}

Here we study what are the most important contributions to the multiwindow cross-correlation signal and test the Limber approximation for these effects.
As we saw that the lensing convergence terms  $\kappa$ are the most relevant relativistic projection effect, we consider the lensing-lensing ($\kappa\times\kappa$) and density-lensing ($\delta_m\times\kappa$) cross-correlations. The performance of Limber approximation is compared to the full computation in Fig.~\ref{fig:limbratio}, where the ratio between the maximum value of the spectra is plotted for increasing $\ell$. The selection function is the same as in the previous section, and the bins are chosen as well separated top-hat functions $W_1\in [0.1,0.4]$ and $W_2\in [0.55,0.95]$. The oscillations are due to the discrete integration over the spherical Bessel functions and the discrete set of $k$ modes over which the integration is performed. Nonetheless, a clear trend emerges. The ratio highly deviates from unity for large angular scales (lower $\ell$s), and in this regime the Limber approximation breaks down as a spherical Bessel function is not well approximated by a Dirac delta function. For small angular scales (higher $\ell$s), the ratio approaches unity in the same-bin correlation case (left panel of Fig.~\ref{fig:limbratio}), retrieving validity of the Limber approximation. On the other hand, in the cross-bin correlation case (right panel of Fig.~\ref{fig:limbratio}), the ratio is consistently lower than the same-bin case. The full computation is crucial in this case, as it preserves terms integrated along the line of sight, which are not well modelled by the Limber approximation. The full computation is necessary to properly capture the signal in this case.
However, a full detectability analysis would require the combination of all GR effects (as done for the full surveys) and a proper estimate of the variability of the results for different magnification bias and survey geometries.
    
In Appendix~\ref{app:MW} we show the 2D visualization of the spherical power spectrum for the auto- and cross-bin correlations for the two bins as above. The full computation shows highly oscillating features that are not present in the Limber result. These are due to the nested integration of the spherical Bessel functions over a finite range (the top-hat function effectively truncates the integration region).

\section{Conclusions}
\label{sec:conclusions}
As galaxy surveys get to probe very large fractions of the sky with high precision measurements, we need to model statistical quantities accurately, dropping many simplifying assumptions and taking into account subtle effects that could be neglected until now.
The spherical Fourier-Bessel formalism is a powerful tool to study the clustering of galaxies in this case, as it is naturally suited to include correlations on large scales and incorporate wide angle and relativistic effects.
It has been shown that the spherical (Fourier-Bessel) power spectrum $\mathcal{S}_\ell(k,k')$ can be used to measure and forecast constraints on cosmological parameters, and its modeling has been studied including a variety of improvements in recent years, and is currently considered a promising observable for forthcoming surveys.

In this work, we have extended this formalism to obtain a full modeling that includes all (linear) relativistic effects without relying on the Limber approximation, radial and cross-bin correlations, and using the full covariance.

Using this formalism we then investigated the importance of these effects. In particular, we looked at how including or neglecting them will affect both accuracy and precision of measurements of the non-Gaussianity parameter $\fnl$, for a SPHEREx-{\it like} survey.
While the precision on measuring the value of $\fnl$ does not considerably change when including or neglecting relativistic corrections, we show that failing to include them would cause to misestimate the inferred value of $\fnl$ by non-negligible amounts.

Our analysis reveals a systematic shift in the best-fit inference of $\fnl$ across all configurations studied. The numerical value of the BFS depends on the survey specifications such as the magnification bias (as cosmic magnification is the largest effect among relativistic corrections) and the galaxy population properties. Nonetheless, the result holds in full generality: the presence of a systematic BFS serves as a crucial warning for future inference analyses, emphasizing the potential impact of GR effects and other systematic sources at both the observational and theoretical level. 
Accurate maps are needed to navigate dangerous waters, and in the same way, accurate theoretical modelings of cosmological observables are crucial to fully exploit the potential of future surveys and provide robust constraints on cosmological parameters. Our result has been derived for a specific case and observable, but this should be considered valid for any approximation taken in the theoretical modeling. SFB is a powerful tool to achieve this goal, as it naturally includes wide-angle and relativistic effects to the theoretical modeling of galaxy clustering.

Furthermore, we test the BFS analysis including only the diagonal terms of the spherical power spectrum, showing that the BFS is still consistently present. Note that the single $\mathcal{B}_{\fnl}$ analysis refers to two theoretical modelings of the same summary statistics of our observable, while comparing diagonal and full analysis would imply connecting the inferred parameters from two different statistical approaches of encoding the same $\mathcal{S}_\ell$ summary. This may be misleading, as it does not refer to the same underlying quantity.

Finally, we have shown how the multiwindow approach can be used to isolate the contribution of specific combinations of GR effects, and how the Limber approximation can be used to target specific regions in $k$ space. We have shown that the full computation is crucial in this case, as it preserves terms integrated along the line of sight, which are not well modeled by the Limber approximation.

We provide a prescription to calculate the cross-bin spectrum. We show that, when correlating different redshift bins, integrated relativistic terms are the only source of correlations.
When measuring cross-bin correlations, we show that the Limber approximation systematically underestimates the spherical spectrum, indicating that our non-flat-sky formalism will be needed.

Therefore, the full sky nondiagonal cross-bin spherical power spectrum allows us to look for a clean signal of relativistic projection effects.

\begin{acknowledgments}
The authors would like to thank Mohamed Yousry Elkhashab, Nicola Bellomo, Sarah Libanore, Francesco Spezzati, Eleonora Vanzan, Sabino Matarrese, and Nicola Bartolo, for useful discussions during the development of this project.
We thank the referee for their insightful comments and suggestions, which significantly enhanced the manuscript.i

This work is partly supported by ICSC–Centro Nazionale di Ricerca in High Performance Computing, Big Data and Quantum Computing, funded by European Union–NextGenerationEU.
D.B. acknowledges partial financial support from the COSMOS
network (www.cosmosnet.it) through the ASI
(Italian Space Agency) Grants 2016-24-H.0, 2016-24-H.1-2018 and
2020-9-HH.
A.R. acknowledges funding from the Italian Ministry of University and Research (MIUR)
through the ``Dipartimenti di eccellenza'' project ``Science of the Universe.''
\end{acknowledgments}

    \appendix

    \section{GENERAL RELATIVISTIC EFFECTS}\label{app:GR}
    \subsection{Definitions and conventions}\label{app:GR_def}
    In this section we provide a brief review of the general relativistic effects that we include in our analysis.

    Recall the expression for the overdensity field defined in Eq.~\eqref{observed_deltag}:
    \begin{equation}
        \delta_{\mathrm{obs}}(\mathbf{n}, z)=\delta_g^{\mathrm{int}}(\mathbf{n}, z)+\delta_{\mathrm{rsd}}(\mathbf{n}, z)+\delta_{\mathrm{v}}(\mathbf{n}, z)+\delta_\kappa(\mathbf{n}, z)+\delta_{\mathrm{pot}}(\mathbf{n}, z)\,.
        \end{equation}
    
The single contributions are defined as follows~\cite{Challinor_2011}:

\begin{align}
    \label{eq:GR_terms}
    \delta_g^{\mathrm{int}}(\mathbf{n}, z)= & b(z) \delta_{v}(r(z) \mathbf{n}, \tau(z)) \\
        \delta_{\mathrm{rsd}}(\mathbf{n}, z)= & \frac{1}{\mathcal{H}(z)} \partial_r(\mathbf{V} \cdot \mathbf{n}) \\
        \delta_{\mathrm{v}}(\mathbf{n}, z)= & {\left[\frac{\mathcal{H}^{\prime}}{\mathcal{H}^2}+\frac{2-5 s(z)}{r \mathcal{H}}+5 s(z)-f_{\mathrm{evo}}(z)\right](\mathbf{V} \cdot \mathbf{n})+\left[3-f_{\mathrm{evo}}(z)\right] \mathcal{H} \Delta^{-1}(\nabla \cdot \mathbf{V}}) \\
        \delta_\kappa(\mathbf{n}, z)= & -(2-5 s(z)) \kappa=-\frac{(2-5 s(z))}{2} \int_0^{r(z)} d r \frac{r(z)-r}{r(z) r} \Delta_2(\Phi+\Psi) \\
        \delta_{\mathrm{pot}}(\mathbf{n}, z)= & (5 s(z)-2) \Phi+\Psi+\mathcal{H}^{-1} \Phi^{\prime} \\
        & +\left[\frac{\mathcal{H}^{\prime}}{\mathcal{H}^2}+\frac{2-5 s}{r \mathcal{H}}+5 s(z)-f_{\mathrm{evo}}(z)\right]\left[\Psi+\int_0^{r(z)} d r\left(\Phi^{\prime}+\Psi^{\prime}\right)\right] \\
        & +\frac{2-5 s}{r(z)} \int_0^{r(z)} d r(\Phi+\Psi)\,.
        \end{align}
Here $\Phi$ and $\Psi$ are the Bardeen potentials,$\mathbf{V}$ is the peculiar velocity, $\delta_v$ is the density contrast in comoving gauge (the off-diagonal component of the energy momentum tensor is null, see next subsection) and $\mathcal{H}$ is the conformal Hubble parameter. Note that the bias $b(z)$ is reported here as scale independent, assuming Gaussianity of the density field. However, when introducing Primordial non-Gaussianity, a scale-dependent term is introduced in the bias function, as discussed in Sec.~\ref{sec:results}.

$r(z)=\tau_0-\tau(z)$ is the conformal distance on the lightcone, and all local quantities are evaluated at conformal time $\tau(z)$, the redshift $z$ and the direction $-\mathbf{n}$, where $\mathbf{n}$ is the direction of the photon propagation path.

\subsection{Evolution and magnification bias}\label{app:magbias}
When considering a galaxy number density distribution $n(z)$ measured by a survey, one has to account for the redshift evolution of such objects. New galaxies form, and the true number density of galaxies does not scale as the inverse cube of the cosmological scale factor $a^{-3}$.  Furthermore, in realistic surveys, additional contributions arise since the galaxy sample is flux limited. The luminosity threshold $L_{\text{min}}$ above which a detection is possible is derived by the observed flux $f_{\mathrm {obs}}$,
\begin{equation}\label{app:lmin}
    L_{\text{min}}=4 \pi \overline{\mathcal{D}}_l^2(z) f_{\mathrm {obs}}\,,
\end{equation}
where the luminosity distance $\overline{\mathcal{D}}_l^2(z)$ is computed in a homogeneous universe. The number density of galaxies that can be detected above the flux limit $f_\mathrm{lim}$ by an observer at $z=0$ is the integral of the luminosity function $\phi(L,z)$ over the luminosity threshold $L_\mathrm{lim}(z)$:
\begin{equation}\label{n_lmin}
    \bar{n}_\mathrm{g}(z)=n(L_\mathrm{lim}(z),z)=\int_{L_\mathrm{lim}(z)}^\infty
    \phi(L,z)\,\dif L\;,
\end{equation}
where $\phi\propto L^{-\gamma}$ with $\gamma>1$ if a constant-slope approximation holds for the differential luminosity function. An additional completeness fraction $0 \leq c_z \leq1$ (here considered constant for simplicity) can also be included to account for the fact that the survey may not pick up all galaxies at a given $z$. However, the logarithmic derivatives do not depend on $c_z$.

We can then define the evolution bias $b_e$ as
\begin{equation}\label{app:bedef}
b_e=-\frac{\partial\ln\bar{n}_g}{\partial\ln (1+z)}\,.
\end{equation}
 A partial logarithmic derivative acts on the mean number density of galaxies with a measured redshift, due to its dependence on both luminosity and redshift $z$~\cite{yousry}. Plugging equation ~\eqref{n_lmin} into Eq.~\eqref{app:bedef} and performing the derivative~\cite{Challinor_2011,Jeong_2012},
\begin{equation}
b_e(z)=-\left.\frac{\partial \ln \int_{L_\text{min}}^\infty
    \phi(L,z)\,\dif L}{\partial \ln (1+z)}
\,
\right|_{L_\text{min}=L_\text{lim}(z)}=-\frac{1}{\bar{n}_\text{g}(z)} \,\int_{L_\text{lim}(z)}^\infty
    \frac{\partial \phi(L,z)}{\partial \ln (1+z)}  \,\dif L  \;.
\end{equation}

One can relate $b_e$ to the derivative of $\Bar{n}_g$ by explicitly writing the derivative and making use of the relation in Eq.~\eqref{n_lmin}~\cite{yousry}:
\begin{equation}
-\frac{\dif \ln \bar{n}_\text{g}(z)}{\dif \ln (1+z)}
=-\frac{1}{\bar{n}_\text{g}(z)}\,\frac{\dif  \bar{n}_\text{g}(z)}{\dif \ln (1+z)}  =2\,Q(z)\,\frac{\dif \ln D_\text{L}(z)}{\dif \ln (1+z)}+b_e(z)\;,
\end{equation}
where $D_\text{L}(z)=(1+z)\, x(z)$ (in a flat universe) so that
\begin{equation}
    \frac{\dif D_\text{L}(z)}{\dif  (1+z)}=x(z)+\frac{c\,(1+z)}{H(z)}\;,
\end{equation}
and $Q(z)$ is the slope of the luminosity function defined as~\cite{Matsubara_2000,magbias1,magbias2,yoo_releff} 
\begin{equation}
Q(z)=-\left.\frac{\partial \ln \bar{n}\left(L_{\min }, z\right)}{\partial \ln L_{\min }}\right|_{L_{\min }=L_{\lim }(z)}= -\frac{L_\text{min}}{\bar{n}(L_\text{min},z)}\,\left. \frac{\partial  \bar{n}(L_\text{min},z)}{\partial L_\text{min}}\right|_{L_\text{min}=L_\text{lim}(z)}=\frac{L_\text{lim}(z)\,\phi(L_\text{lim}(z),z)}{\bar{n}_\text{g}(z)}\,.
\end{equation}

Finally, the full relation reads~\cite{Bertacca_2015_magbias}
\begin{equation}
 -\frac{\dif \ln \bar{n}_\text{g}(z)}{\dif \ln (1+z)}=2\,Q(z)\,\left[1+ \frac{c\,(1+z)}{H(z)\,x(z)} \right]+b_e(z)\,.
 \end{equation}

Assuming again a constant-slope approximation for the differential luminosity function ($\phi\propto L^{-\gamma}$ with $\gamma\equiv$ cost), the coefficient $Q$ can be rewritten as $Q=(\gamma-1)$, which allows us to relate it to the slope $s$ expressed in terms of magnitude $M=\mathrm{cost}-2.5\log_{10}(L/L_0)$,
\begin{equation}
    s=\diff{\log_{10}\bar{n}_g(M)}{M}\,,
\end{equation}
so that $s=\frac{2}{5}(\gamma-1)$ and $Q=\frac{5}{2}s$.

\section{SFB FORMALISM}\label{app:coeff}
In this appendix, we provide more details concerning the SFB basis and the overdensity field expansion. We also address some of the numerical challenges of this work.

First, we write the plane wave representation as
\begin{equation}\label{planewave}
\braket{\mathbf{x}}{\mathbf{k}}=\sum_{\ell m}(4 \pi) i^{\ell} j_{\ell}(k r) Y_{\ell m}^*(\hat{\mathbf{k}}) Y_{\ell m}(\hat{\mathbf{x}})\,,
\end{equation}
where $\mathbf{x}=r\hat{\mathbf{x}}$ is the position vector and $\mathbf{k}=k\hat{\mathbf{k}}$ encodes the Fourier mode, with $r=\abs{\mathbf{x}}$, $k=\abs{\mathbf{k}}$ and $\hat{\mathbf{x}}$, $\hat{\mathbf{k}}$ unit directional vectors.
The $\ket{k \ell m}$ basis has been defined in Eq.~\eqref{eq:klm_def}. The orthonormality of the basis then implies that
\begin{gather}
\braket{k\ell m}{k'\ell'm'} =1 \delta^D(k-k') \delta_{\ell \ell'} \delta_{m m'} \; , \\
\braket{\mathbf{k}'}{k \ell m}=\frac{(2 \pi)^{3 / 2}(-i)^{\ell}}{k} Y_{\ell m}(\hat{\mathbf{k}}) \delta^D(k-k')\,.
\end{gather}

Furthermore, the completeness relation is defined as
\begin{equation}\label{completeness}
    1=\int d^3\mathbf{x}\ket{\mathbf{x}}\bra{\mathbf{x}}=\int \frac{d^3\mathbf{k}}{(2\pi)^3}\ket{\mathbf{k}}\bra{\mathbf{k}}=\sum_{\ell m}\int dk\ket{k \ell m}\bra{k \ell m}\,.
\end{equation}

Thus, with this prescription, any scalar field $\varphi=\ket{\varphi}$ can be projected on $\ket{\mathbf{x}}$, $\ket{\mathbf{k}}$ and $\ket{k\ell m}$ through
\begin{gather}
    \braket{\mathbf{x}}{\varphi}=\varphi(\mathbf{x})\\
    \braket{\mathbf{k}}{\varphi}=\varphi(\mathbf{k})\\
    \braket{k \ell m}{\varphi}=\varphi_{\ell m}(k)\,.
\end{gather}
Using the completeness conditions~~\eqref{completeness}, the projected fields can be written as
\begin{align}
    \varphi(\mathbf{x})&=\int \frac{d^3\mathbf{k}}{(2\pi)^3}\braket{\mathbf{x}}{\mathbf{k}}\braket{\mathbf{k}}{\varphi}\,,\\
    \varphi(\mathbf{k})&=\int d^3\mathbf{x}\braket{\mathbf{k}}{\mathbf{x}}\braket{\mathbf{x}}{\varphi}\, \\
        \varphi_{\ell m}(k)&=\int d^3 \mathbf{x}\braket{k \ell m}{\mathbf{x}}\braket{\mathbf{x}}{\varphi}=\int \frac{d^3 \tilde{\mathbf{k}}}{(2 \pi)^3}\braket{k \ell m}{\tilde{\mathbf{k}}}\braket{\tilde{\mathbf{k}}}{\varphi}\,,
\end{align}
where two equivalent formulations are provided for $\varphi_{\ell m}(k)$.

For a generic scalar filed $\varphi$, the spherical power spectrum $\mathcal{S}_\ell(k,k')$ is defined as the ensemble average of the product of the spherical Fourier modes of the field, $\varphi_{\ell m}(k)$ and its complex conjugate $\varphi_{\ell' m'}^*(k')$. That is~\cite{rassat,YD13},
\begin{equation}
\langle\varphi_{\ell m}(k) \varphi_{\ell' m'}^*(k')\rangle \equiv \delta_{\ell \ell'} \delta_{m m'} \mathcal{S}_\ell(k, k')\,,
\end{equation}
where $\delta_{\ell \ell'}$ and $\delta_{m m'}$ are Kronecker delta functions.

The spherical Fourier mode $\varphi_{\ell m}(k)$ can then be expressed in terms of the field $\varphi(\mathbf{x})$ as
\begin{equation}
\varphi_{\ell m}(k)=\int d^3 \mathbf{x} \sqrt{\frac{2}{\pi}} k j_\ell(k r) Y_{\ell m}^*(\hat{\mathbf{x}}) \varphi(\mathbf{x})
\end{equation}
Here, $j_\ell(k r)$ is the spherical Bessel function of order $\ell$, $Y_{\ell m}^*(\hat{\mathbf{x}})$ is the complex conjugate of the spherical harmonic function of degree $\ell$ and order $m$, and $\hat{\mathbf{x}}$ is the unit vector in the direction of $\mathbf{x}$.

This leads to an expression for the spherical power spectrum:
\begin{eqnarray}\label{sph_powsp_gen}
\mathcal{S}_\ell(k, k')= \frac{2 k k'}{\pi} \int d^3 \mathbf{x}_1 \int d^3 \mathbf{x}_2
Y_{\ell m}^*(\hat{\mathbf{x}}_1) Y_{\ell m}(\hat{\mathbf{x}}_2) j_\ell(k r_1) j_\ell(k' r_2)\langle\varphi(\mathbf{x}_1) \varphi(\mathbf{x}_2)\rangle \, .
\end{eqnarray}

\subsection{Two-point function}
\label{app:2pt}
The two-point correlation function~\cite{Feldman_1994,YD13} is derived for the overdensity field to compute its power spectrum. Using Eq.~\eqref{sph_powsp_gen} to expand the number density field, the correlation between overdensity Fourier modes $\delta_{\ell m}(k)$ reads :
\begin{equation}\label{sl_2point}
\begin{aligned}
\langle\delta_{\ell m}(k) \delta_{\ell' m'}^*\qty(k')\rangle \equiv & \frac{2}{\pi} k k' \int d r_1 r_1^2 W(r_1) \int d r_2 r_2^2 W(r_2)j_\ell\qty(k r_1) j_{\ell'}\qty(k' r_2) \\
&\times\int d^2 \hat{\mathbf{x}}_1 \int d^2 \hat{\mathbf{x}}_2 Y_{\ell m}^*\qty(\hat{\mathbf{x}}_1) Y_{\ell' m'}\qty(\hat{\mathbf{x}}_2)\frac{\langle n(\mathbf{x}_1) n\qty(\mathbf{x}_2)\rangle }{\tilde{n}_g^2}\,,
\end{aligned}
\end{equation}
with~\cite{Feldman_1994}
\begin{equation}\label{expansion}
   \frac{\langle n(\mathbf{x}_1) n\qty(\mathbf{x}_2)\rangle }{\tilde{n}_g^2}=\Big\{\phi\qty(r_1) \phi\qty(r_2)\qty(1+\xi_g\qty(\mathbf{x}_1-\mathbf{x}_2))+\frac{1}{\tilde{n}_g} \phi\qty(r_1) \delta^D\qty(\mathbf{x}_1-\mathbf{x}_2)\Big\}\,.
\end{equation}

The two-point correlation function can be related to the power spectrum of the comoving curvature perturbation $\zeta(\mathbf{k})$ as
\begin{equation}\label{xifour}
\xi_g\qty(\mathbf{x}_1-\mathbf{x}_2)=\int \frac{d^3 k}{(2 \pi)^3} \mathcal{T}_g\qty(k, r_1) \mathcal{T}_g\qty(k, r_2) P_{\zeta}(k) e^{i \mathbf{k} \cdot\qty(\mathbf{x}_1-\mathbf{x}_2)}\,,
\end{equation}
where a transfer function $\mathcal{T}_g$ connects the power spectrum to the standard power spectrum of primordial curvature perturbation $\zeta$.

Eq.~\eqref{xifour} can be inserted into expansion~\eqref{sl_2point}, using Eq.~\eqref{planewave} to simplify the integration over the angular part, which involves the product of spherical harmonics and plane waves. The orthogonality condition for the spherical harmonics allows us to reduce the double angular integral in the combination of Eqs.~\eqref{sl_2point} and~\eqref{xifour} to
\begin{equation}\label{spharm_step}
\int d^2 \hat{\mathbf{x}}_1 \int d^2 \hat{\mathbf{x}}_2 Y_{\ell m}^*\qty(\hat{\mathbf{x}}_1) Y_{\ell' m'}\qty(\hat{\mathbf{x}}_2) e^{i \mathbf{k} \cdot\qty(\mathbf{x}_1-\mathbf{x}_2)}=(4 \pi)^2 i^{\ell-\ell'} j_\ell\qty(k r_1) j_{\ell'}\qty(k r_2) Y_{\ell m}^*(\hat{\mathbf{k}}) Y_{\ell' m'}(\hat{\mathbf{k}})\,.
\end{equation}

The $\xi_g$-dependent term in Eq.~\eqref{sl_2point} then becomes:
\begin{equation}
\begin{aligned}
\mathcal{\bar{S}}_\ell\qty(k, k') \delta_{\ell \ell'} \delta_{m m'}=&
32 \pi k k' i^{\ell-\ell'} \int d r_1 r_1^2 \int d r_2 r_2^2 W(r_1)\phi(r_1) W(r_2)\phi(r_2) j_\ell(k r_1) j_{\ell'}(k' r_2) \\&
\times\int \frac{d^3 \tilde{k}}{(2 \pi)^3} \mathcal{T}_g(\tilde{k}, r_1) \mathcal{T}_g(\tilde{k}, r_2) j_\ell(\tilde{k} r_1) j_{\ell'}(\tilde{k} r_2) Y_{\ell m}^*(\hat{\tilde{\mathbf{k}}}) Y_{\ell' m'}(\hat{\tilde{\mathbf{k}}}) P_{\zeta}(\tilde{k}) \\
=&(4 \pi) \int d \ln \tilde{k} \Delta_{\zeta}(\tilde{k})\Bigg[\frac{2}{\pi} k k' \int d r_1 r_1^2 W(r_1)\phi(r_1) j_\ell(k r_1) j_\ell(\tilde{k} r_1) \mathcal{T}_g(\tilde{k}, r_1) \\
&\times \int d r_2 r_2^2 W(r_2)\phi(r_2) j_\ell(k r_2) j_\ell(\tilde{k} r_2) \mathcal{T}_g(\tilde{k}, r_2)\Bigg] \delta_{\ell \ell'} \delta_{m m'} \,.
\end{aligned}
\end{equation}
The expression can be made more compact by introducing the spherical multipole function $\mathcal{M}_\ell(\tilde{k}, k)$, that is
\begin{equation}\label{sphmult_app}
\mathcal{M}_\ell(\tilde{k}, k) \equiv k \sqrt{\frac{2}{\pi}} \int_0^{\infty} d r r^2 W(r)\phi(r) j_\ell(\tilde{k} r) j_\ell(k r) \mathcal{T}_g(\tilde{k}, r)\,,
\end{equation}
which finally leads to
\begin{equation}\label{signal_def_app}
\mathcal{\bar{S}}_\ell(k, k')=4 \pi \int d \ln \tilde{k} \Delta_{\zeta}(\tilde{k}) \mathcal{M}_\ell(\tilde{k}, k) \mathcal{M}_\ell(\tilde{k}, k')\,.
\end{equation}

The shot-noise part in Eq.~\eqref{sl_2point} can be derived by decomposing the 3D Dirac delta function as
\begin{equation}
    \delta^D(\mathbf{x}_1-\mathbf{x}_2)=\frac{\delta^D(r_1-r_2)\delta^D(\hat{\mathbf{x}}_1-\hat{\mathbf{x}}_2)}{r_1^2}\,.
\end{equation}
The noise component then reads
\begin{equation}
\begin{aligned}
\mathcal{N}_\ell\qty(k, k') \delta_{\ell \ell'} \delta_{m m'}&=\frac{2 k k'}{\pi \tilde{n}_g} \int d r_1 r_1^2 W(r_1) \int d r_2  r_2^2 W(r_2)\phi(r_2) j_\ell(k r_1) j_{\ell'}(k' r_2)\frac{ \delta^D(r_1-r_2)}{r_1^2}\\
&\times\int d^2 \hat{\mathbf{x}}_1 Y_{\ell m}^*\qty(\hat{\mathbf{x}}_1) Y_{\ell' m'}\qty(\hat{\mathbf{x}}_1)\,,
\end{aligned}
\end{equation}
and it can be rewritten more compactly as:
\begin{equation}\label{noise_def_app}
\mathcal{N}_\ell\qty(k, k') \equiv \frac{2 k k'}{\pi \tilde{n}_g} \int_0^{\infty} r^2 W^2(r)\phi(r) j_\ell(k r) j_\ell\qty(k' r)\,d r\,.
\end{equation}

In sum, the spherical Fourier power spectrum is the sum of the two contributions,
\begin{equation}
\mathcal{S}_\ell\qty(k, k') \equiv \bar{\mathcal{S}}_\ell\qty(k, k')+\mathcal{N}_\ell\qty(k, k')\,.
\end{equation}

\subsection{Numerical integration}\label{app:numint}
The highly oscillating nature of the spherical Bessel functions make the numerical integration of the spherical power spectrum~\eqref{sph_mult_win} a challenging task.  This is a known issue in the literature, with several methods developed to tackle the issue. In cosmological applications, a Fast Fourier Transform FFTlog approach to double-spherical Bessel integrals has been proposed in~\cite{fftlog}, which exploits a power-law expansion of the integrand to retrieve a sum of terms that can be solved analytically. Furthermore,~\cite{2fast} built upon the method by providing a pipeline to perform accurate computations of of one- and two- spherical Bessel function integrals. 

However, the structure of the spherical multipoles in Eq.~\eqref{sph_mult_win}, combined in the full power spectrum equation as in Eq.~\eqref{signal_def}, makes it difficult to apply the methods in~\cite{fftlog,2fast} directly. The issue arises from the fact that the window function introduce a finite range of integration and integrated effects show a nontrivial dependence on the line-of-sight distance. A generalization of such methods would be required to efficiently introduce a FFTlog-like approach to the problem.

To overcome this issue and to limit the computational cost, we employ Gauss-Legendre quadrature, that allows for fast and accurate integration over a finite range. The method is defined for a integration range $[-1,1]$ and is based on the discretization of the integral to a weighted sum evaluated at specific nodes $x_i$: 
 \begin{equation}
    \int_{-1}^{1}  f(x)dx  \approx \sum_{i=1}^{n}  w_{i}  f(  x_{i} )\,.
\end{equation}

Any finite bounded interval $[a,b]$ can be properly mapped to the $[-1,1]$ interval by means of a change of variable $x=t(b-a)/2+(a+b)/2$. The integrations limits that are formally semi-infinite can be mapped to a finite range by introducing a cutoff $r_{\max}$ and $\tilde{k}_{\max}$. $r_{\max}$ has a clear physical interpretation as the maximum comoving distance of the observable Universe (note however that any realistic selection function kills any contribution well before these scales). $\tilde{k}_{\max}$ has been fixed through a convergence test, increasing its value until the spherical power spectrum does not change within a given tolerance. The method has been tested for a range of multipoles and for different survey specifications, showing a good convergence for the spherical power spectrum.

\subsection{Power spectrum discretization and boundary conditions}
\label{app:kdisc}
Note that throughout this work, the SFB decomposition has been performed assuming $r_{\rm min}=0$ and full sky coverage. Dropping the assumption of a (theoretically) infinite survey (i.e., defined by its selection function at each redshift), one can impose an additional boundary condition at some $r_{\rm max}=R$ constraining the total volume over which the SFB transform is performed [$\delta(r>R)=0$].
In order to keep the basis function orthogonal, a discrete spectrum of modes $k=k_{n,\ell}$ arises for each angular mode $\ell$~\cite{fish95}.
For such a finite survey, the orthogonality condition now reads
    \begin{equation}
        \int_0^R \mathrm{~d} r r^2 j_\ell(k r) j_\ell\qty(k^{\prime} r)=\delta_{k k^{\prime}}^K C_{\ell n}^{-1}\,,
        \end{equation}

and it holds if the radial wave numbers satisfy
    \begin{equation}
        A j_\ell^{\prime}(k R)=B \frac{j_\ell(k R)}{k R}\,.
        \end{equation}
$A,B$ are arbitrary constants that are picked so that the solution is phisically well motivated.

Following~\cite{fish95}, neglecting fluctuations outside $R$ requires to match the solution of Poisson's equation for $r<R$ and Laplace's equation for $r\geq R$ . One must then impose continuity of the field and its logarithmic derivative at $r=R$. This results in a condition for $k_n$:
    \begin{equation}
        j_\ell'\qty(k_n R)+(\ell+1) j_\ell\qty(k_n R) / k_n R=0\,,
        \end{equation}
    which can be reduced to
    \begin{equation}
        j_{\ell-1}\qty(k_n R)=0\,.
    \end{equation}
The discrete spectrum of radial wave numbers is then given by the zeros of the spherical Bessel function of order $\ell-1$.

\begin{figure}
        \centering
        \includegraphics[width=0.5\textwidth]{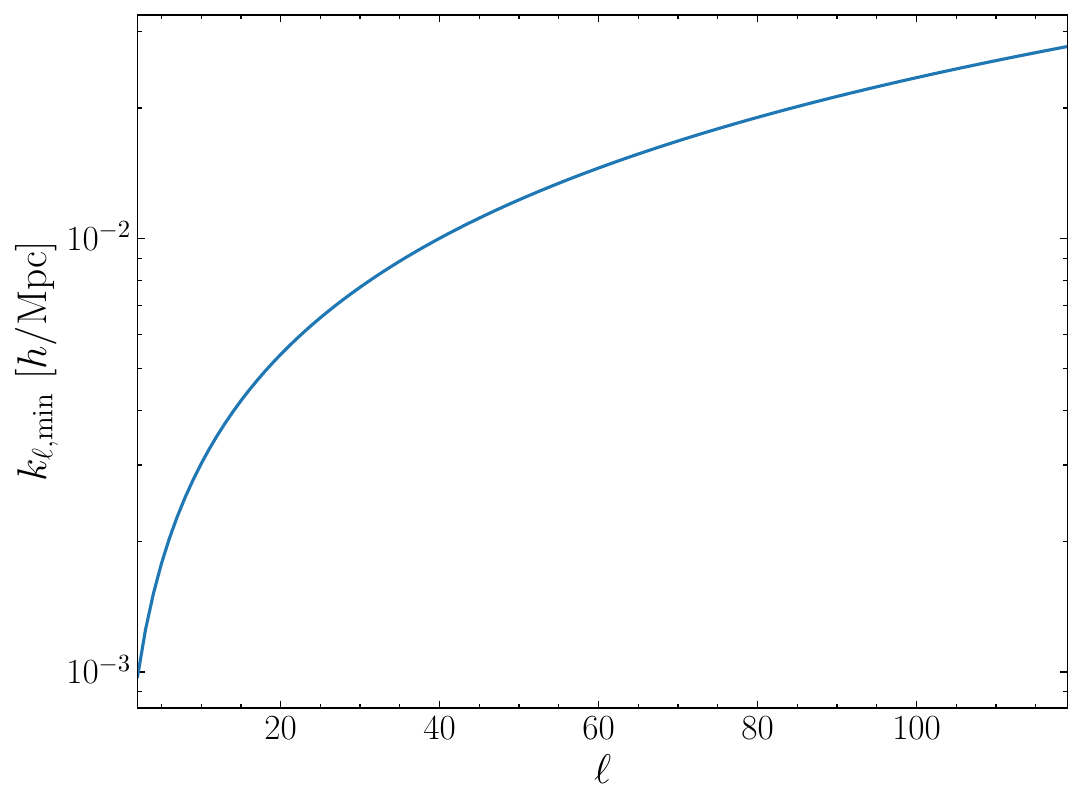}
        \caption{\justifying Minimum $k_{\min}(\ell)$ for each $\ell$ for the survey depth $R=4609.6 \,{\rm Mpc/h}$, which corresponds to $z_{\max}\sim 3.2$, which is compatible with a SPHEREx-like survey.}
        \label{fig:kmin}
\end{figure}

In Fig.~\ref{fig:kmin}, we plot the minimum $k_{\min}(\ell)$ for each $\ell$ for the survey depth $R=4609.6 \,{\rm Mpc/h}$, which corresponds to $z_{\max}\sim 3.2$. This value of $R$ is derived by the SPHEREx 4 sample used in the main analysis and from the volume computation defined in Eq.~\eqref{defrmax}.

\section{BEST-FIT SHIFTS}

\subsection{Bias marginalization and contour plots}
\label{app:contour}
\begin{figure}
    \centering
    \includegraphics[width=\textwidth]{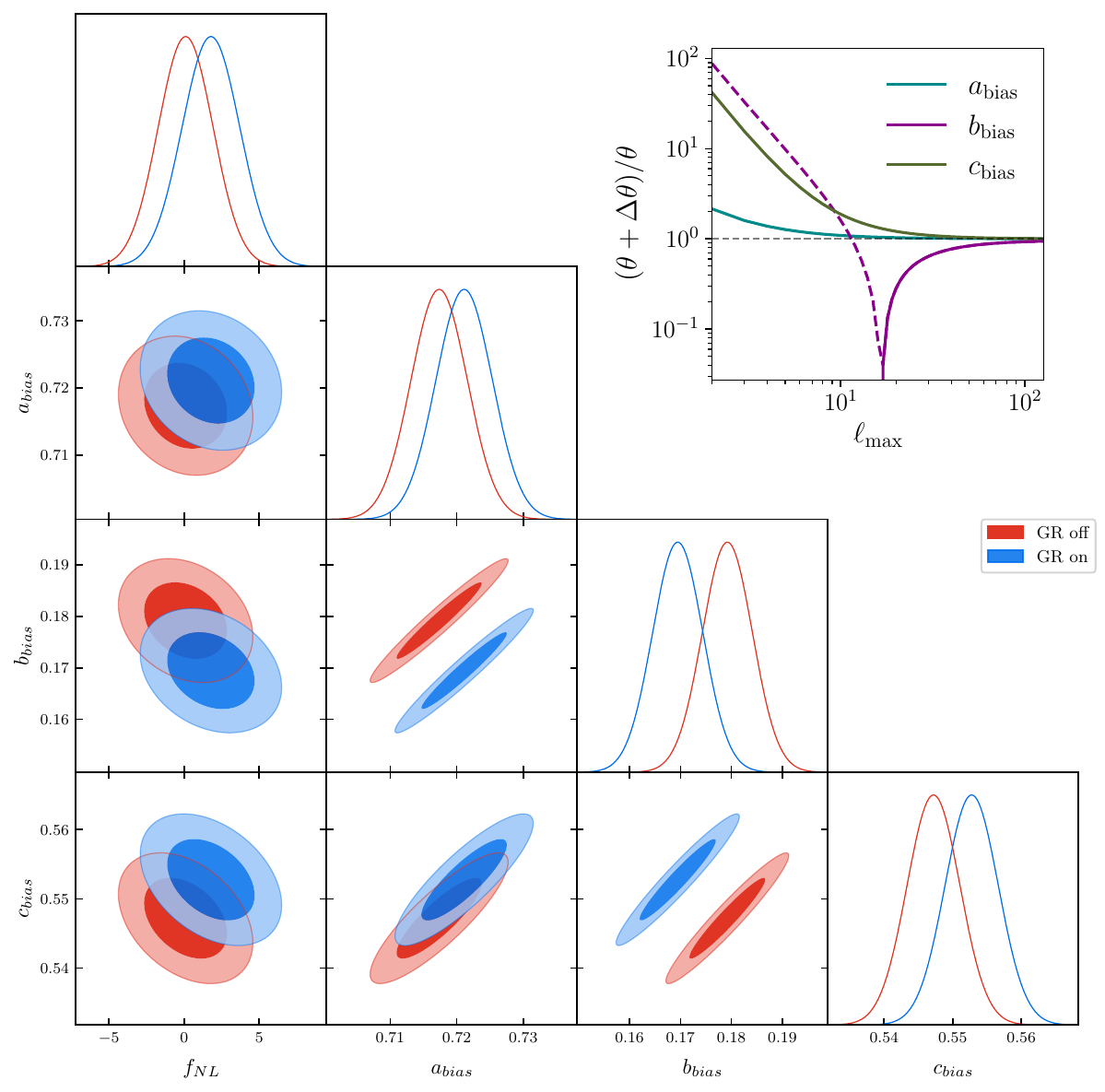}
    \caption{\justifying Marginalized corner plot for the diagonal Fisher matrix analysis at $\ell_{\max}=128$  and $s=0.6$. Spurious shift on bias parameters is suppressed more slowly with respect to the off-diagonal case.  Recall that the bias function is fitted as $b(x,a_{\mathrm{bias}},b_{\mathrm{bias}},c_{\mathrm{bias}})=a_{\mathrm{bias}}(1+b_{\mathrm{bias}}\cdot z(x))^{c_{\mathrm{bias}}}$. See main text for details.}
    \label{fig:marg1d}
\end{figure}

\begin{figure}
    \centering
    \includegraphics[width=\textwidth]{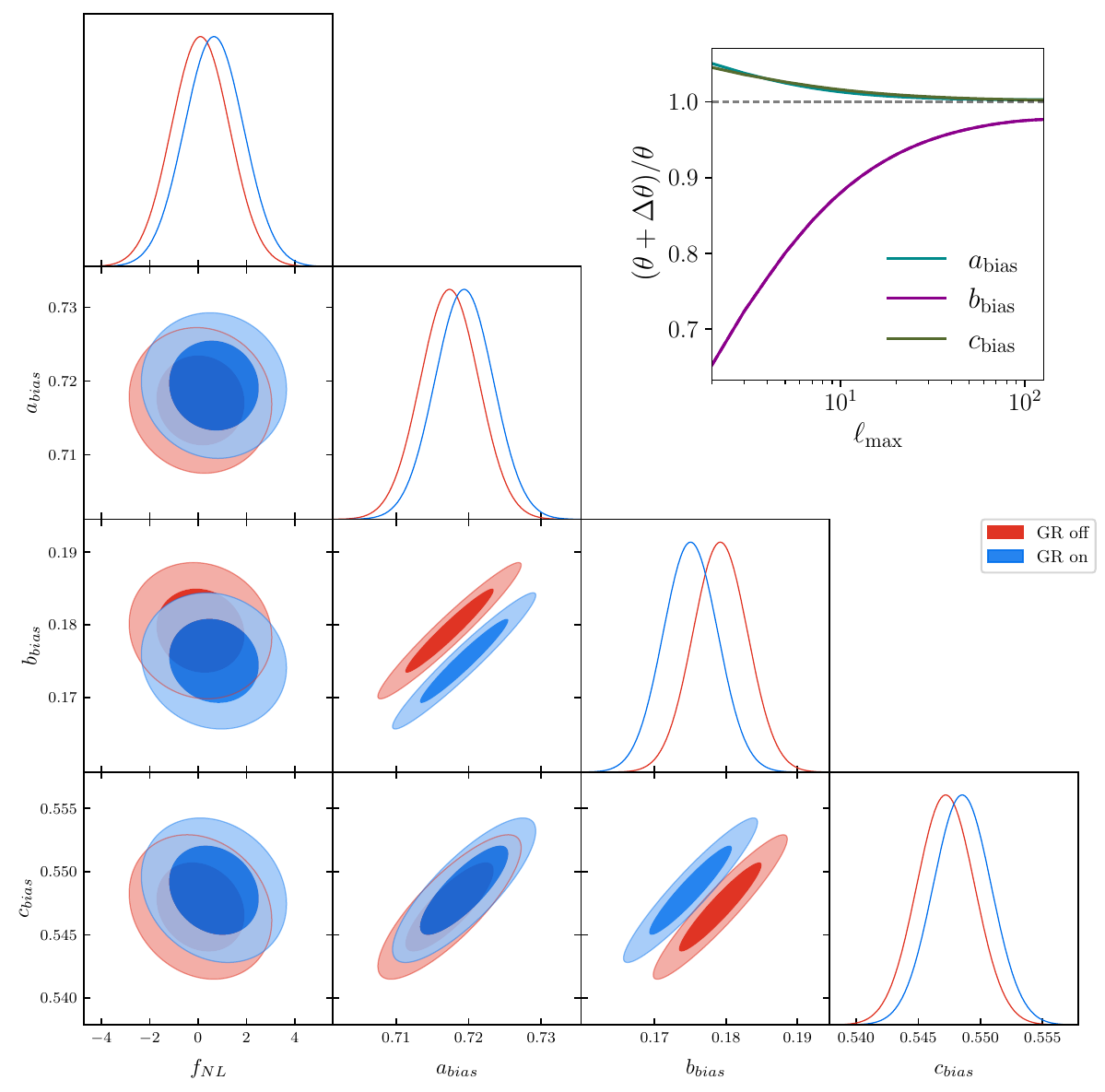}
    \caption{\justifying Same as Fig.~\ref{fig:marg1d} but for the full analysis. Marginalised corner plot for the full Fisher matrix analysis at $\ell_{\max}=128$  and $s=0.6$.}
    \label{fig:marg2d}
\end{figure}

To better visualize the impact of BFSs over parameter constraints, 2D contour plots are shown both for the diagonal analysis in Fig.~\ref{fig:marg1d} and full analysis in Fig.~\ref{fig:marg2d}. As reported in the main text, the shift on the bias parameters significantly reduces in both approaches, although convergence is slower for the diagonal analysis and the final ratio is still relevant, making the respective confidence ellipses separated in parameter space. Both approaches preserve a relevant BFS for the $\fnl$ parameter.

Both figures also show the ratio of the shifted bias parameters with respect to its fiducial value for increasing $\ell_{\max}$, both for the diagonal and full analysis. The shift $\Delta\theta$ (with $\theta$ any of the three bias parameters $a_{\mathrm{bias}},b_{\mathrm{bias}},c_{\mathrm{bias}}$) is effectively suppressed when including higher $\ell$s. However, in the diagonal analysis the cumulative shift reduces at higher $\ell_{\rm max}$ with respect to the full case: this is due to the fewer modes available for each angular scale. Specifically, at each angular scale $\ell$, the diagonal analysis has  $n_k^{ \ell }$ $k$ modes, while the full analysis includes $n_k^{ \ell }(n_k^{ \ell }+1)/2$ modes i.e.,~the upper-diagonal part of the $\mathcal{S}_\ell(k,k')$ matrix to avoid double counting.

Although the BFS is significantly reduced by including smaller angular scales, it is still relevant when compared to the expected $1\sigma$ precision.


\subsection{Diagonal analysis}
\label{app:diagonal}
\begin{figure}
    \centerline{
    \includegraphics[width=\textwidth]{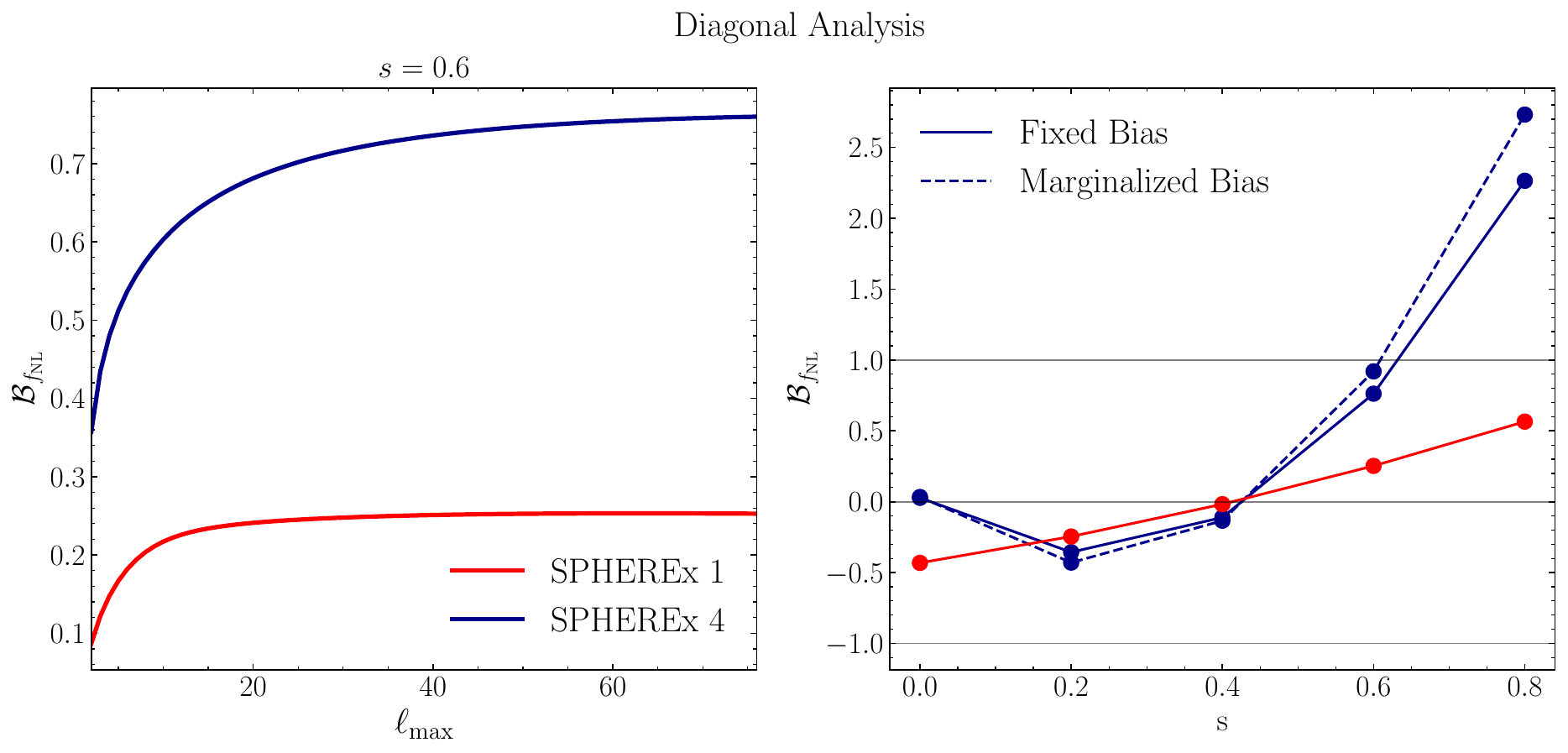}}
    \caption{\justifying Same as Figure\ref{fig:bfs2d} but for the diagonal analysis, using just the $\mathcal{S}_\ell(k,k)$ terms of the spherical power spectrum. Left panel: the curves trace the ratio $\mathcal{B}_{\fnl}\equiv\Delta \fnl / \sigma_{\fnl}$. Right panel: the ratio $\mathcal{B}_{\fnl}$ for a set of constant magnification bias values $s$.}
    \label{fig:bfs2d_diag}
    \end{figure}

In Fig.~\ref{fig:bfs2d_diag}, we show the BFS for the diagonal analysis, where only the $\mathcal{S}_\ell(k,k)$ terms of the spherical power spectrum are used. The results are consistent with the full analysis, showing a relevant BFS for the $\fnl$ parameter consistently for all the configuration tested. Again, for $s=0.4$ the lensing convergence term $\kappa$ is suppressed and $\mathcal{B}_{\fnl}\sim 0$.

\section{SPHERICAL MULTIPOLES IN THE LIMBER APPROXIMATION}\label{app:MW}
Here we derive the expressions for the spherical multipoles $\mathcal{M}_\ell^{\delta,i}(\tilde{k}, k)$ in the Limber approximation, defined in Sec.~\ref{sec:limber}.

In the Limber approximation, one can compute the integral of a generic function $f(r)$ convolved with a pair of spherical Bessel functions of order $\ell$ and $\ell'$ as
\begin{equation}
\int dr\, r^2 f(k,\tilde{k},r) j_\ell(kr)j_{\ell^{'}}(\tilde{k}r) \sim \frac{\pi}{2}\frac{1}{k^2}\sqrt{\frac{\nu}{\nu'}}f\qty(k,\frac{\nu'}{\nu}k,\frac{\nu}{k})\delta^D\qty(\tilde{k}-\frac{\nu'}{\nu}k)\,.
\end{equation}
    
This allows us to simplify the spherical multipole expression for the matter fluctuation $\delta_m$ and lensing convergence $\kappa$ contributions, they read~\cite{YD13}
    \begin{gather}
    \mathcal{M}_\ell^{\delta,i}(\tilde{k}, k) = \sqrt{\frac{\pi}{2 k^2}} W_{i}\qty(\frac{\nu}{k}) \mathcal{T}_m\qty(k, \frac{\nu}{k}) \delta^D(k-\tilde{k})\,,\\
    \mathcal{M}_\ell^{\mathcal{K},i}(\tilde{k}, k) = \sqrt{\frac{\pi}{2 \tilde{k}^2}} W_{i}\qty(\frac{\nu}{k}) \frac{3 H_0^2}{2} \Omega_m \mathcal{T}_m\qty(\tilde{k}, \frac{\nu}{\tilde{k}})\ell(\ell+1)\qty(\frac{\tilde{k}-k}{\tilde{a} k^2 \tilde{k}^2})\,.\label{limblens}
    \end{gather}

    The Kaiser RSD term includes the second derivative of the product of the transfer function $\mathcal{T}(k,r)$ and a spherical Bessel function. The derivative can be connected to spherical Bessel functions of order $\ell-1$ and $\ell+1$. We finally get (where for simplicity $\mathcal{A}_i(r)\equiv W_i(r)\phi(r)f(r)$) 
    \begin{align}
        \mathcal{M}_\ell^{R S D,i}(\tilde{k}, k) \simeq &  \sqrt{\frac{\pi}{2}}\frac{1}{k}\mathcal{A}\qty(\frac{\nu}{k})\qty(\mathcal{T}^{''}\qty(k,\frac{\nu}{k})  - \frac{2(\ell+1)}{\nu}\mathcal{T}^{'}\qty(k,\frac{\nu}{k}) + \frac{\ell^2-\ell-\nu^2}{\nu^2}\mathcal{T}\qty(k,\frac{\nu}{k}))\delta^D(k-\Tilde{k})\\
        &+\sqrt{2\pi}\frac{1}{k}\sqrt{\frac{\nu}{\nu'}}\mathcal{A}\qty(\frac{\nu}{k})\mathcal{T}^{'}\qty(\frac{\nu'}{\nu}k,\frac{\nu}{k})\delta^D\qty(\tilde{k}-\frac{\nu'}{\nu}k)\\
        &+\sqrt{2\pi}\frac{1}{k}\sqrt{\frac{1}{\nu\nu''}}\mathcal{A}\qty(\frac{\nu}{k})\mathcal{T}\qty(\frac{\nu''}{\nu}k,\frac{\nu}{k})\delta^D\qty(\tilde{k}-\frac{\nu''}{\nu}k)
    \end{align}
    with $\nu'=\ell-1/2$ and $\nu''=\ell+3/2$.

    While the window function product $W_1(\nu / k)W_2(\nu / k')$ acts on $(k,k')$, the integration in $\tilde{k}$ still spans all possible values. This may lead to issues since redshift-dependent quantities are computed as a function of $z(\nu/\Tilde{k})$, which becomes nonphysical for low $\Tilde{k}$. We overcome this problem by considering the relation
    \begin{equation}
         k=\frac{\nu}{r} = f(a)\,,
    \end{equation}
    where $f$ is a generic function that connects $k$ to the scale factor $a$ (in practice this is implemented via interpolation). This allows us to rewrite the expression by performing a change of variables:
    \begin{equation}
    \mathcal{S}_\ell(k, k')=4 \pi \int_0^1 d a \frac{df(a)}{da}\Delta(f(a)) \mathcal{M}_l(f(a), k) \mathcal{M}_l(f(a), k')\,,
    \end{equation}
    which ensures a well-defined integral.

\section{MULTIWINDOW POWER SPECTRUM}

\begin{figure}
    \centering
    \includegraphics[width=0.75\linewidth]{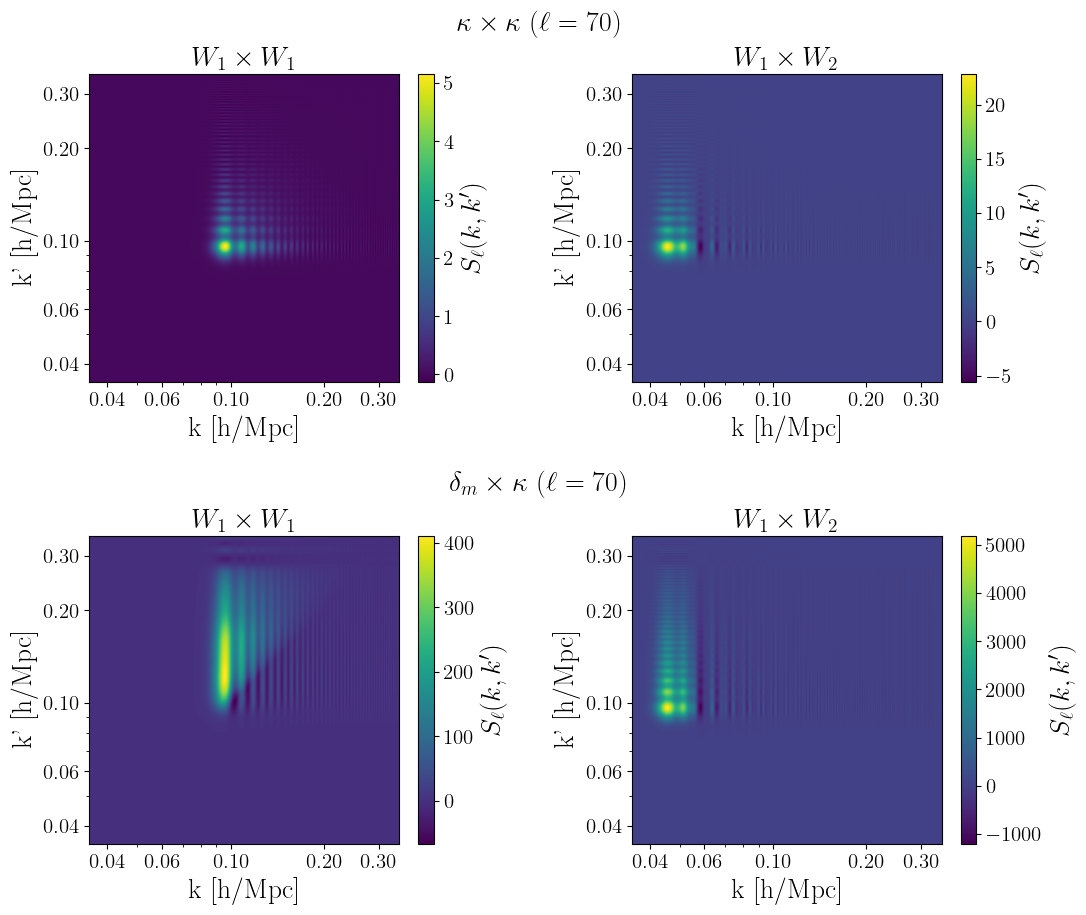}
    \caption{\justifying 2D visualization of the cross-correlation for two bins that span respectively the redshift range $W_1\in [0.1,0.4]$ and $W_2\in [0.55,0.95]$. Both autobin and cross-bin correlation is showed for integrated effects and terms involving mixed local-integrated effects.
    Top panels: lensing convergence $\kappa$ terms.
    Bottom panels: local matter overdensity $\delta_m$ and Lensing convergence $\kappa$ mixed term.
    The cross-correlations are localized nondiagonally while autocorrelations are placed along the diagonal, as expected. The oscillations are due to the top-hat window functions that effectively make the integration region for the spherical multipoles $\mathcal{M}_{\ell}(\tilde{k}, k)$ finite.}
    \label{fig:mwsurv}
    \end{figure}

The full power spectrum can be computed by combining the contributions from the different windows. Here we show some of the most relevant terms for the cross-correlation of two bins $W_1$ and $W_2$, which span, respectively, the redshift ranges $[0.1,0.4]$ and $[0.55,0.95]$. The top panels of Fig.~\ref{fig:mwsurv} show the autobin and cross-bin correlation for the lensing convergence $\kappa$ terms, while the bottom panels show the local matter overdensity $\delta_m$ and lensing convergence $\kappa$ mixed term. One can indeed see how for high $\ell$ the relevant nonzero contributions are mainly localized in the areas depicted in Fig.~\ref{fig:mwvis}. The relevant information is contained in the cross-bin correlation, which is more sensitive to the integrated effects. The mixed term is also relevant, as it contains information on the local matter overdensity and the lensing convergence. As discussed in the main text, the oscillations in all the spectra are due to the finite integration interval in the spherical multipoles computation $\mathcal{M}_{\ell}(\tilde{k}, k)$, as a consequence of the top-hat window functions.

\begin{figure}
    \centering
    \includegraphics[width=\linewidth]{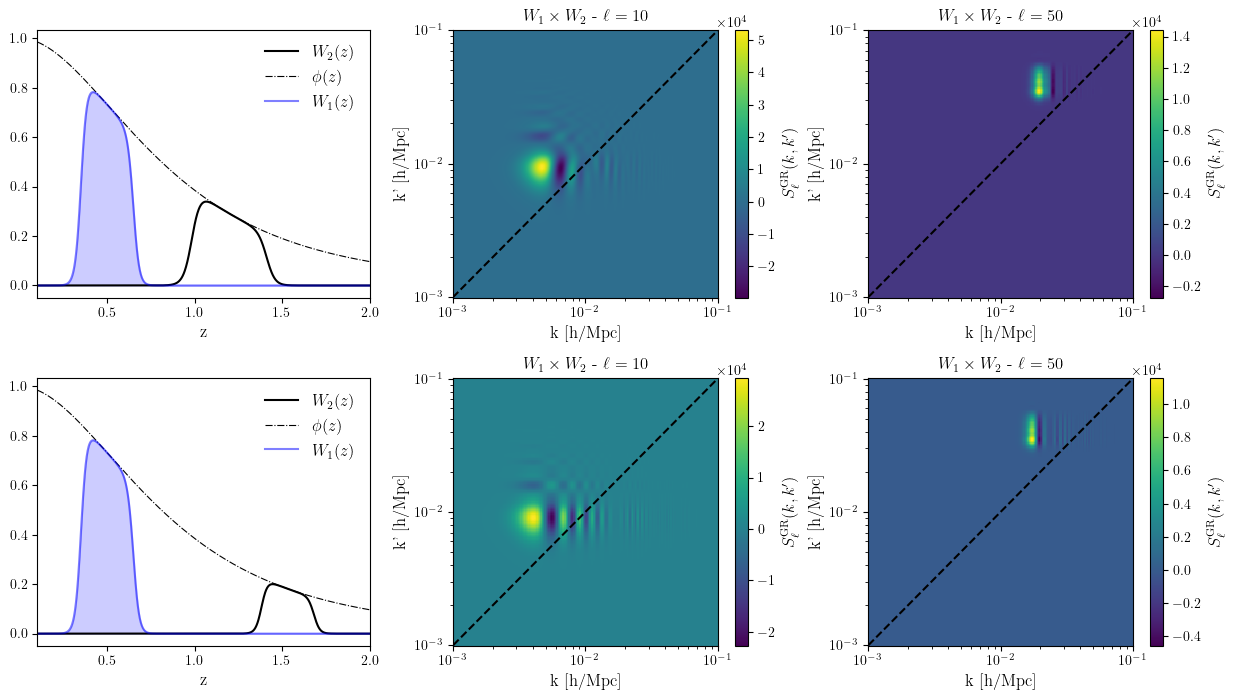}
    \caption{\justifying Full cross-bin spherical power spectrum for two window functions with
smoothed edges. The correlation peaks off-diagonally due to the presence of integrated effects that dominate for separated bins, as discussed in the main text. Similarly to Fig.~\ref{fig:mwsurv}, the oscillating features are still present due to the discrete integrations
introduced by the window functions.}
    \label{fig:mwtanh}
    \end{figure}

Fig.~\ref{fig:mwtanh} shows the full cross-bin spherical power spectrum for two pairs of bins at different redshift distance and smoothed edges. One can see that the oscillating features are still present in the power spectrum due to the discreteness introduced by the window functions. The cross-bin correlation is localized nondiagonally, as expected.

\bibliographystyle{unsrtnat}
\bibliography{filtered_bellomized}
\end{document}